\documentclass[reprint,notitlepage,twocolumn,
superscriptaddress,aps,longbibliography,nofootinbib]{revtex4-1}
\usepackage[utf8]{inputenc}
\usepackage{braket}
\usepackage{amsmath}
\usepackage{amsthm}
\usepackage{mathrsfs}
\usepackage{mathtools} 
\usepackage{amssymb}
\usepackage{dsfont}
\usepackage{braket}
\usepackage{bm}
\DeclareMathOperator{\tr}{tr}
\usepackage{color}
\pdfoutput=1

\usepackage[english]{babel}
\usepackage[T1]{fontenc}
\usepackage{amsmath}
\usepackage[colorlinks=true]{hyperref}
\usepackage{enumitem}
\usepackage{amsthm}
\usepackage{amssymb}
\usepackage{graphicx}
\usepackage{float}
\usepackage[ruled]{algorithm2e}

\begin{document}

\title{
Noise-Agnostic Quantum Error Mitigation with Data Augmented Neural Models
}
\author{Manwen Liao}
\affiliation{QICI Quantum Information and Computation Initiative, Department of Computer Science,
The University of Hong Kong, Pokfulam Road, Hong Kong}
\author{Yan Zhu}
\thanks{Manwen Liao and Yan Zhu contribute equally}
\affiliation{QICI Quantum Information and Computation Initiative, Department of Computer Science,
The University of Hong Kong, Pokfulam Road, Hong Kong}

\author{Giulio Chiribella}
\affiliation{QICI Quantum Information and Computation Initiative, Department of Computer Science,
The University of Hong Kong, Pokfulam Road, Hong Kong}
\affiliation{Department of Computer Science, Parks Road, Oxford, OX1 3QD, United Kingdom}
\affiliation{Perimeter Institute for Theoretical Physics, Waterloo, Ontario N2L 2Y5, Canada}

\author{Yuxiang Yang }
\email{yuxiang@cs.hku.hk}
\affiliation{QICI Quantum Information and Computation Initiative, Department of Computer Science,
The University of Hong Kong, Pokfulam Road, Hong Kong}

\begin{abstract}
    Quantum error mitigation, a data processing technique for recovering the statistics of target  processes from their noisy version, is a crucial task for near-term quantum  technologies.   Most existing methods require prior knowledge of the noise model or the noise parameters. Deep neural networks have a potential to lift this requirement,  but current models require training data  produced by ideal processes in the absence of noise.
Here we build a neural model that achieves quantum error mitigation  without any prior knowledge of the noise and without training on noise-free data. To achieve this feature, we introduce a quantum augmentation technique  for  error mitigation.   Our approach applies to quantum circuits and to the dynamics of many-body and continuous-variable quantum systems, accommodating various types of noise models.  We demonstrate its effectiveness by testing it both on simulated noisy circuits and on real quantum hardware.

\end{abstract}

\maketitle

\section{Introduction}
Quantum technologies have potentially disruptive applications in the fields of  computing, communication, and sensing. In the near term, however,  the demonstration of practical quantum advantages   remains challenging due to the presence of noise.
A promising technique to restore quantum advantages in realistic noisy devices is  quantum error mitigation~\cite{cai2022quantum}, including zero-noise extrapolation (ZNE)~\cite{temme2017error,li2017efficient, kandala2019error, kim2023scalable, kim2023evidence}, Clifford data regression (CDR)~\cite{czarnik2021error, czarnik2022improving, strikis2021learning, Lowe_2021}, probablistic error cancellation~\cite{temme2017error, endo2018practical, van2023probabilistic}, and virtual purification~\cite{PhysRevX.11.041036, PhysRevX.11.031057, PRXQuantum.4.010303}.


A limitation of the existing error mitigation methods is that they generally require prior knowledge about the noise model, leading to an overhead in terms of noise characterization operations~\cite{harper2021fast, harper2020efficient,figueroa2021randomized}. 
A promising approach to circumvent this issue is to exploit deep neural networks,  which  have been successfully applied to other quantum tasks such as   quantum state characterization~\cite{torlai2018neural, zhu2022flexible, ahmed2021quantum, carrasquilla2019reconstructing, du2023shadownet}, quantum property estimation~\cite{wu2023learning,tang2023q}, quantum verification~\cite{wu2023quantum, qian2023multimodal}, and quantum simulations~\cite{sharir2020deep,bennewitz2022neural}. Previous research~\cite{zhang2021direct, qin2024experimental, vadali2024quantum} has also explored the use of machine learning models to estimate the output fidelity of quantum circuits. Although these approaches do not eliminate noise in the circuit output, they hold promise in guiding the generation of quantum circuits with fewer errors. Recently, a series of works explored the application of deep neural networks directly to quantum error mitigation~\cite{kim2020quantum, liao2023machine, Zhukov_2022, sack2024large}. Nevertheless, training these networks generally requires access to noise-free data, which can be hard to obtain from experiments  or from classical simulations.  

Here  we propose a neural model that achieves quantum error mitigation without prior knowledge of the noise and without any access to noise-free data. To achieve this feature,    we  introduce  a technique, called quantum data augmentation,  to expand the original data set by generating new data from a fiducial set of noisy processes. Our technique provides a quantum version of classical data augmentation techniques~\cite{shorten2019survey, taylor2018improving},  which have proven valuable in scenarios where the available training data is limited \cite{wang2020generalizing,wang2019survey}.


Our  model exhibits four major features.   (1) No need of noise-free statistics from the target quantum process. Thanks to this feature, our model is applicable  to relevant real-world scenarios where the ideal target process is hard to simulate.  (2) Noise-agnostic error mitigation. The model does not require prior knowledge about the noise model, nor about the values of the noise parameter.  As a result, it avoids  overheads due to noise characterization, and works both for Markovian and non-Markovian types of noise. 
(3) Versatility. The model works in a broad range of applications and enables error mitigation for quantum algorithms, dynamics of many-body systems, and continuous-variable quantum information processing. In addition,  it accepts the input data in a variety of different forms, including expectation values of quantum observables, statistics of measurement outcomes, and estimates of the Wigner function. (4) Transferability. The trained model exhibits the capability to mitigate errors for circuits sharing the same circuit skeleton as those considered in the training, all without the need for retraining. This feature makes it possible to apply the model to a wide range of quantum circuits, enhancing its practical utility and scalability.
To demonstrate these features, we test our model on a series of paradigmatic quantum algorithms, such as variational quantum eigensolvers~\cite{peruzzo2014variational} and quantum approximate optimization~\cite{farhi2014quantum}, and quantum dynamics, such as the many-body dynamics of the Ising model and the Kerr Hamiltonian~\cite{dykman2012} for  continuous variable systems. Furthermore, we tested our model on real quantum hardware. The results demonstrate its superior performance compared to previous methods, including ZNE and CDR.

\section{Results}
\label{sec:results}

\subsection{The DAEM model} Let us start by specifying our error mitigation framework.  
We consider a target quantum process  $\cal E$ corresponding to a quantum circuit composed of a specific sequence of single-qubit and CNOT gates (which is general as single-qubit gates and CNOT gates can form a universal gate set), with the restriction that only Pauli measurements are performed. The circuit represents the action of an ideal quantum device in the absence of noise. Note that our method also applies to some reversible processes without explicit circuit representations (see Section \ref{sec:cv}).
In the real world, however, one has only access to noisy versions of the process $\cal E$.    Such noisy versions will be denoted by  $\mathcal{N}_{\lambda}(\mathcal{E})$, where $\mathcal{N}_{\lambda}$ represents the noise model and ${\lambda}$ indicates the noise parameter. The input state  of the process $\mathcal{N}_{\lambda}(\mathcal{E})$  is randomly selected from an ensemble $\mathcal{S}=  \{  \rho_s\}_{s=1}^n$, which can generally contain multiple quantum states.  For every input state $\rho_s  \in  \mathcal{S}$, the goal of error mitigation  is to estimate the statistics of measurements performed on the ideal output state ${\cal E}  (\rho_s)$ given access to data from its noisy version $\mathcal{N}_{\lambda}(\mathcal{E})  (\rho_s)$.

Here we focus on a set of Pauli measurements $\mathcal{M}$ of interest, such as a set of Pauli measurements performed on a subset of the output qubits in a quantum computation. 
Each measurement $\boldsymbol{M}_i = (M_{ij})_j$ in $\mathcal{M}$ is a positive operator-valued measure (POVM) consisting of $m$ positive operators that satisfy the normalization condition $\sum_{j=1}^{m} M_{ij} = \mathds{1}$.  This general setup covers tomography (when $\mathcal{M}$ is informationally complete), as well as quantum algorithms, where a single measurement  is used for read-out.  


\begin{figure*}[hbtp]
    \centering
    \includegraphics[width=0.6\textwidth]{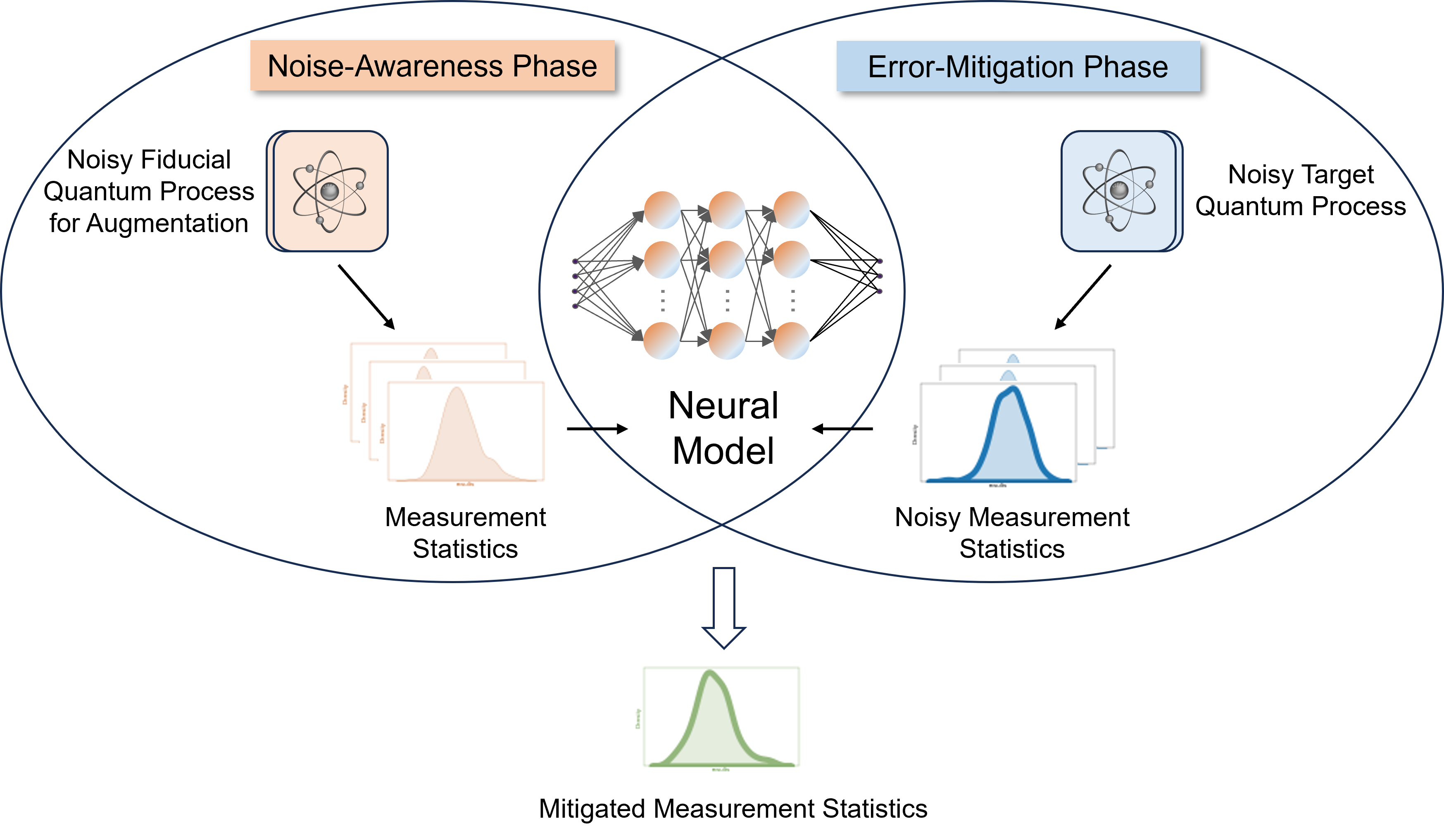}
    \caption{ Framework of DAEM model. The entire procedure is divided into two phases. In the first phase, known as Noise-Awareness phase, we train the neural model for error mitigation with the assistance of a fiducial process responsible for data augmentation.
    In the second phase, known as Error-Mitigation phase, we apply the trained neural model to mitigate the errors in noisy measurement statistics collected from the noisy version of the target quantum process. }
    \label{fig:framework}
\end{figure*}

To achieve error mitigation,  we now introduce a neural network-based model. Our model, called   data augmentation-empowered error mitigation model (DAEM), is illustrated in Figure~\ref{fig:framework}.   Its high-level structure consists of two phases. In the first phase, the Noise-Awareness phase, we train a neural network to remove the action of the noise ${\cal N}_\lambda$ from the measurement statistics.  The training is boosted by a technique called {\em quantum data augmentation}. {The key idea is to train the network on data generated by fiducial processes, a set of quantum circuits derived from the target quantum process.}

{The fiducial process $\mathcal F$ is expected to have two desired features: 
First, in the absence of noise, a classical computer should be capable of efficiently generating adequate measurement data corresponding to various input states and Pauli measurements. In the presence of noise, the fiducial process ${\cal N}_\lambda ({\cal F})$ should be implementable using the same quantum computing hardware that executes the target process.
Second, when implemented on the quantum hardware, the noise pattern of the fiducial process should be as close to that of the target process as possible. This ensures that error mitigation techniques learned on the fiducial process can be effectively transferred to the target process. }

Following this spirit, we construct the fiducial process by making changes to the execution of the original implementation according to the following recipe: (1) {For every single-qubit gate $R$, we instead ask the quantum computer to execute $\sqrt{R^\dagger} \sqrt{R}$, which equals an identity gate in the ideal case. The motivation for this is to make the noise pattern of the fiducial process emulates that of the target process. For example, in trapped ion systems, the replaced gates can be implemented with the same execution time by adjusting the duration of the interaction. This leads to similar noise patterns assuming the dissipative part of the qubit dynamics to be fixed.}
Note that the implemented fiducial process will not be an identity process in general, since the implementation is not perfect. 
(2) All CNOT gates are executed according to the original circuit. 

{With this recipe, since ${\cal F}$ consists only of CNOT gates, the measurement statistics for the output state ${\cal F} (\sigma)$ with respect to Pauli measurements can be efficiently computed for any product state $\sigma$: Note that ${\cal F}$ is Clifford since it consists of CNOTs only. Taking the Heisenberg picture, the evolution of any Pauli observable $P$ under ${\cal F}^{\dag}$, which results in an $N$-qubit Pauli observable ${\cal F}^{\dag}(P)$, can be efficiently simulated. Since $\sigma$ is a product state, the desired expectation $\tr({\cal F}^{\dag}(P)\sigma)$ can be computed classically in $O(N)$ time.

}


 { To generate the training data, the experimenter collects measurement statistics by executing the noisy fiducial processes $\mathcal{N}_{\lambda}(\mathcal{F})$ on a set of product states $\{\sigma_s\}$, using the same hardware as the target quantum process. The  acquired measurement statistics will be denoted by $\{{\bm{p}'}_{i,s}^{(1)}\}$, while the ideal measurement statistics will be denoted by  ${\bm{p}'}_{i,s}^{(0)}:=  \big(\tr(\mathcal{F}(\sigma_s)M_{ij})\big)_j$.  In scenarios where varying the noise is possible, statistics can also be collected with various noise parameters $\{\lambda_k\}_{k=1}^{K}$. In such cases, the acquired measurement statistics are denoted by $\{{\bm{p}'}_{i,s}^{(k)}\}_{k=1}^{K}$. Note that we do not assume knowledge of the exact values of $\{\lambda_k\}_{k=1}^{K}$ and, consequently, do not require any extra noise estimation procedure.  }
 We  then train the neural model by providing it tuples of the form $( {\bm{p}'}_{i,s}^{(k)} )_{k=1}^K$ corresponding to a given input state $\sigma_{s}$ and a given measurement $  \bm{M}_i$.  Thanks to the aforementioned features of the fiducial processes, the ideal statistics ${\bm{p}'}_{i,s}^{(0)}$  can be computed efficiently from the input product state $\sigma_{s}$.   In the training, we optimize the parameters of the model with respect to  a loss function $\mathcal{L}$ that quantifies the deviation between the predicted statistics and the noise-free one (see Methods for details).   

After the training is concluded, the model can be used for error mitigation on the target process $\cal E$.    The experimenter  collects measurement statistics by performing the noisy process $\mathcal{N}_{\lambda}(\mathcal{E})$ on  an arbitrary input state $\rho \in \mathcal{S}$, with  different  noise parameters $\{\lambda_k\}_{k=1}^{K}$.  The corresponding statistics will be denoted by $\boldsymbol{p}_{i}^{(k)}  = \big(\tr(  {\cal N}_{\lambda_k}  (\mathcal{E})(\rho) \,M_{ij})\big)_j$. The neural model then outputs the inferred the ideal statistics  $\boldsymbol{p}_{i}^{(0)}:=  \big(\tr(\mathcal{E}(\rho)M_{ij})\big)_j$ pertaining the target process $\mathcal{E}$. A detailed description of the implementation of the neural model in various examples is provided in the Methods section.

An important feature of the DAEM model is that it can be applied to ensembles containing multiple input states. In addition,  the states appearing in the Error Mitigation phase do not need to be the same states used in the training.  
Furthermore, it is worth stressing that the model  does not require any ideal  measurement data (neither experimentally generated nor classically simulated) for the target process $\mathcal{E}$.   { As a consequence, it has the potential to be applied  to large-scale systems   where the classical simulations are not feasible, and realistic experiments are affected by non-negligible amounts of noise.  }

Additionally, our  model can be trained for multiple target processes that share the same circuit skeleton but have different parameters. This is achieved using a set of fiducial processes. The underlying intuition is that circuits with the same skeleton are likely affected by similar noise patterns. Consequently, the knowledge gained from mitigating errors in one such circuit can be transferred to others within the same structural framework. This transferability enhances the efficiency and applicability of our model, reducing the need for extensive retraining for each new set of parameters.

\begin{figure*}[t]
    \centering
    \includegraphics[width=0.8\textwidth]{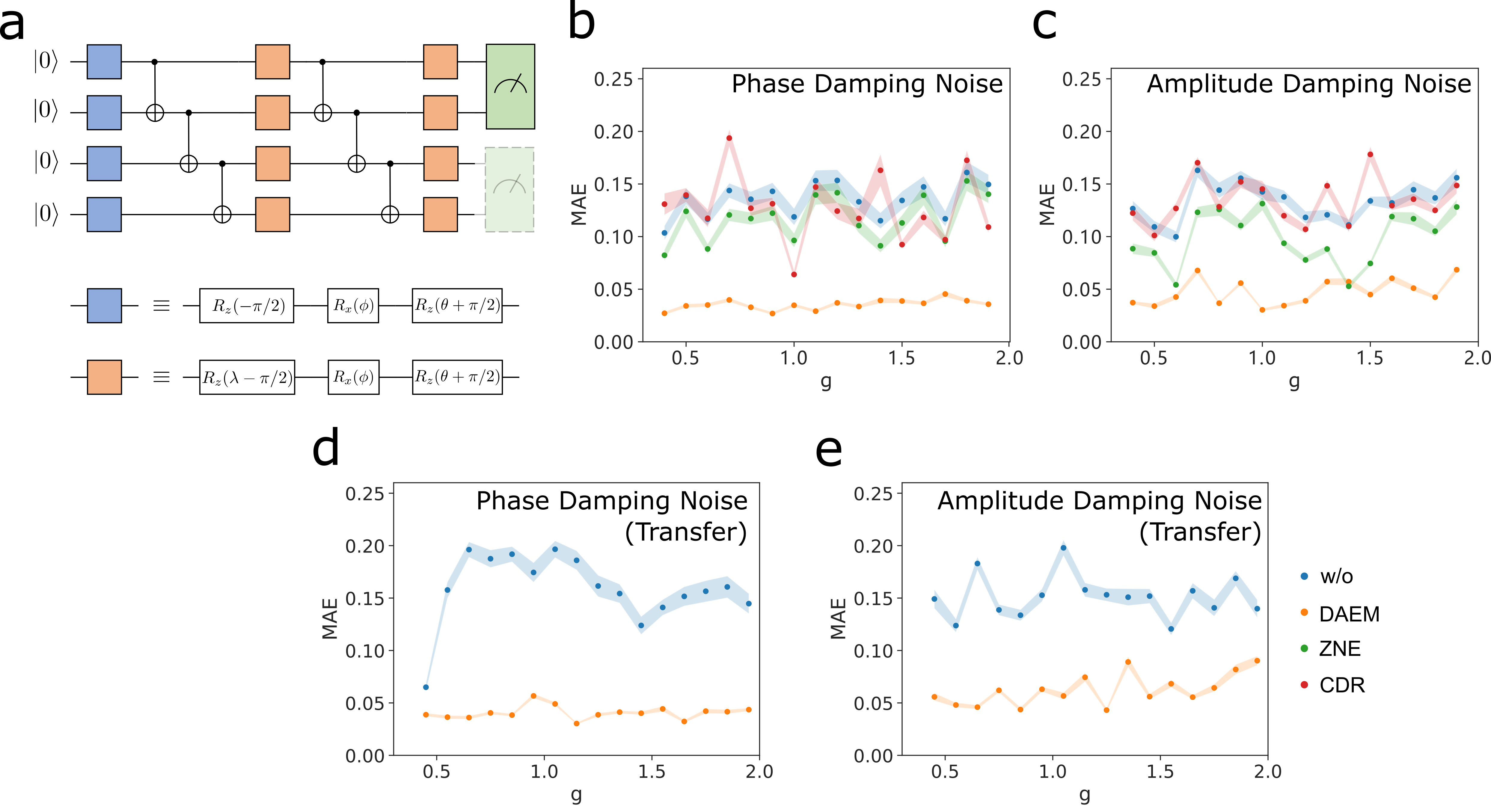}
    
    \caption{Error mitigation for variational quantum eigensolvers. 
    a.\ The variational ansatz for preparing the ground states of 4-qubit transverse Ising models. 
    b.\ Mean Absolute Errors (MAE) between the mitigated measurement expectation values for phase damping noise model and ideal expectation values.
    c.\ Mean Absolute Errors (MAE) between the mitigated measurement expectation values for amplitude damping noise model and ideal expectation values.
     d.\ Mean Absolute Errors (MAE) between the mitigated measurement expectation values for phase damping noise model and ideal expectation values for the circuits not included in the training.
    e.\ Mean Absolute Errors (MAE) between the mitigated measurement expectation values for amplitude damping noise model and ideal expectation values for the circuits not included in the training.  
    It is noteworthy that ZNE requires knowledge of noise parameters associated with statistics while our proposed DAEM not. Despite  being under an unfair comparison, DAEM still demonstrates superior performance.
    }
    
   %

    \label{fig:VQE}
\end{figure*}

\begin{figure}[h]
    \centering
    \includegraphics[width=0.4\textwidth]{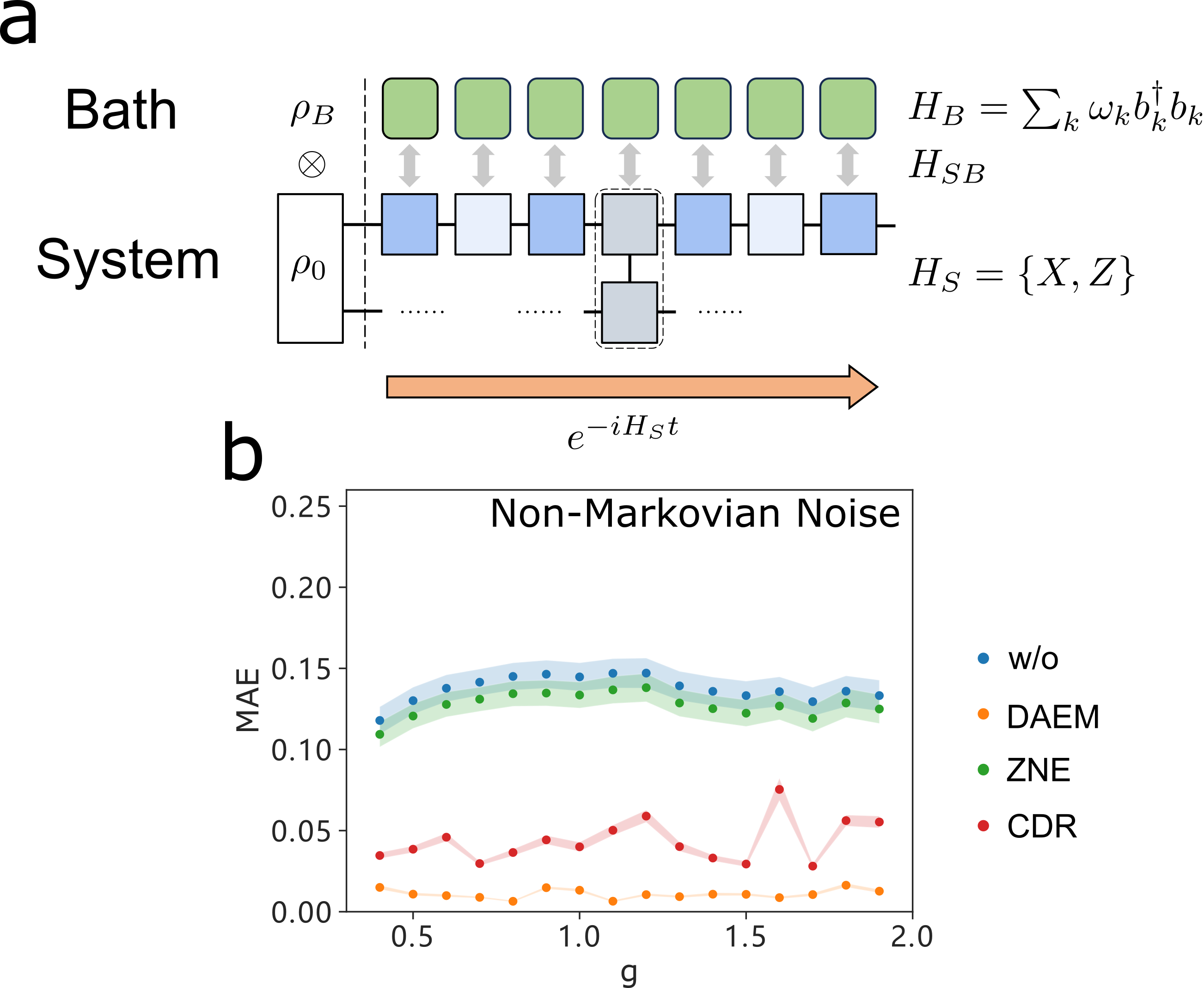}

    \caption{Error mitigation for variational quantum eigensolvers affected by Non-Markovian Noise. 
    a.\  Schematic diagram of quantum circuits affected by Non-Markovian noise. 
    b.\ Mean Absolute Errors (MAE) between the mitigated measurement expectation values for considered Non-Markovian noise model and ideal expectation values. }

    \label{fig:VQE_nm}
\end{figure}

\begin{figure}[hbtp]
    \centering
    \includegraphics[width=0.4\textwidth]{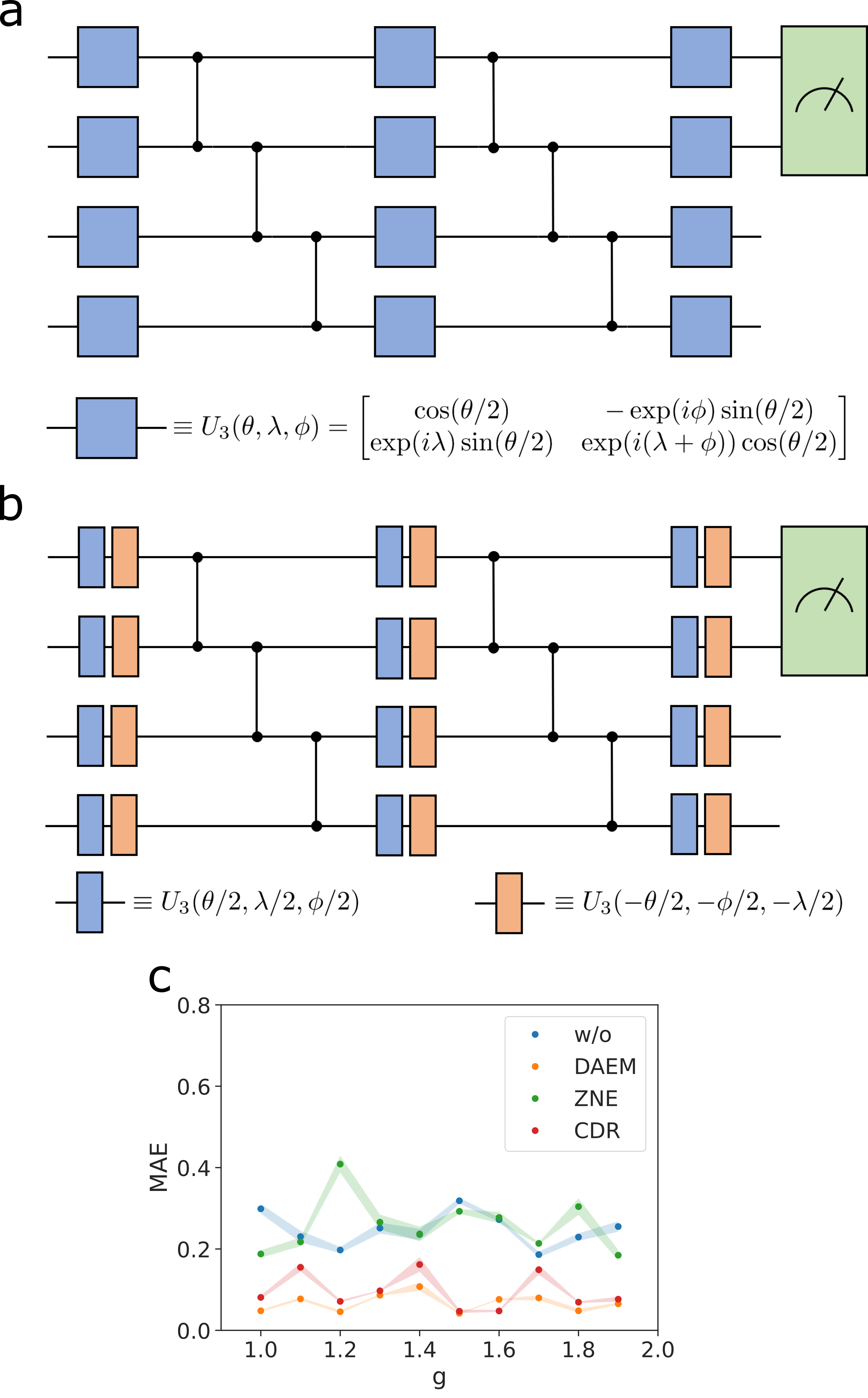}
    
    \caption{ Error mitigation for variational quantum eigensolvers on the OriginQ Cloud  quantum hardware. 
    a.\ The variational ansatz for preparing the ground states of 4-qubit transverse Ising models. 
    b.\ The structure of fiducial circuits.
    c.\ Mean Absolute Errors (MAE) between the mitigated measurement expectation values and ideal expectation values. The average MAE of w/o, DAEM, ZNE, CDR are 0.247, 0.067, 0.259, and 0.095 respectively.
    }
    \label{fig:Vqe_real}
\end{figure}

\subsection{Error Mitigation for Quantum Algorithms}\label{subsec:circuit}
The domain most suitable for testing our error mitigation model is quantum circuits, which are widely employed in various quantum algorithms. Our framework applies generally to quantum algorithms, where the goal is to obtain noise-free statistics from noisy quantum circuits.
In this section, we test the performance of DAEM on prototypical NISQ algorithms, including the Variational Quantum Eigensolvers (VQEs)~\cite{peruzzo2014variational}, the swap test~\cite{buhrman2001quantum}, and the Quantum Approximate Optimization Algorithm (QAOA)~\cite{farhi2014quantum}.

\paragraph{Variational quantum eigensolvers.}  Variational Quantum Eigensolvers (VQEs), widely utilized in the realms of quantum chemistry and quantum computation, leverage parameterized quantum circuits to approximate the ground states of specified Hamiltonians. However, in practical scenarios, these circuits inevitably grapple with noise, leading to deviations in the ground state energy from the ideal scenario. In this context, we consider a scenario where an experimenter possesses the optimal parameters of a well-trained VQE circuit and intends to employ it on a real noisy quantum device. The experimenter's goal is to derive the ideal measurement statistics of the ground state based on the gathered noisy measurement data. 

In the following, we consider the VQEs for the transverse Ising chain with Hamiltonian
\begin{equation}\label{eqn:Ising}
    H_{\text{Ising}} = -g \sum_{i = 1}^N X_i - J\sum_{i = 1}^{N - 1} Z_i Z_{i + 1},
\end{equation}
where $X$, $Z$ are Pauli operators, and $N$ is the number of qubits. The variational ansatz used to prepare the ground state is a hardware-efficient ansatz, composed of single-qubit Euler rotation gates and CNOT gates, as illustrated in Figure~\ref{fig:VQE}a. We choose $16$ circuits, varying the parameter $g$ within the range of $[0.4, 2.0)$ with a stride of $0.1$. Additionally, we set the values of $J$ and $N$ to be $1$ and $4$ respectively for all experiments. For the set of measurements $\mathcal{M}$, we choose all two-qubit Pauli measurements on nearest-neighbor qubits.  In the Noise-Awareness phase, we construct the fiducial circuit by replacing each single-qubit rotation gate with two single-qubit rotation gates, while keeping the CNOT gates unchanged. These rotation gates are parameterized to mutually cancel each other out. 
{  For instance, an $R_x(\phi)$ gate is replaced by an identity gate, which is specifically constructed as $R_x(-\phi/2)R_x(\phi/2)$. }
We let $\mathcal{S}$ be all of the $4$-qubit mixed states and we randomly select $n=100$ states $\{\sigma_{s_1}\}_{s_1=1}^n$ in the Noise-Awareness phase of all our experiments for each $g$.

During the Error-Mitigation phase, we evaluate our mitigation model by using the prepared initial state $\rho_0 = \ket{0}\bra{0}^{\otimes N}$.

First, we evaluate our model's performance under two Markovian noise models: amplitude damping and phase damping. In all of the experiments, noise is applied after each gate in Figure~\ref{fig:VQE}a. The amplitude damping noise channel and the phase damping noise channel are mathematically defined by Equation~\ref{eqn:phasedamping} and Equation~\ref{eqn:amplitudedamping}, respectively. Throughout all of our experiments, we consider a set of noise parameters denoted as $\{\lambda_k\}_{k=1}^K \in [0.05, 0.29]$, with stride 0.02.

\begin{equation}\label{eqn:amplitudedamping}
    \rho \rightarrow V_0\rho V_0^\dag + V_1\rho V_1^\dag, 
\end{equation}
with $V_0 = \begin{bmatrix} 1& 0 \\ 0 & \sqrt{1-\lambda} \end{bmatrix}$ and $V_1 = \begin{bmatrix} 0 & \sqrt{\lambda} \\ 0 & 0  \end{bmatrix}$.
Figure~\ref{fig:VQE}b illustrates the mitigation results obtained using various error mitigation techniques for VQE circuits affected by phase damping noise, while Figure~\ref{fig:VQE}c presents the mitigation results for VQE circuits affected by amplitude damping noise. The results clearly demonstrate that DAEM consistently outperforms other mitigation methods for each VQE circuit, regardless of the specific value of $g$.

Furthermore, we tested our trained model on circuits for preparing ground states of the Ising model with parameters not included in the training set, using the same variational ansatz. Specifically, we varied the parameter $g$ within the range of [0.45, 1.95] with a stride of 0.1. We present the experimental results for mitigating phase damping noise and amplitude damping noise in Figure~\ref{fig:VQE}d and Figure~\ref{fig:VQE}e. The results demonstrate that our model can efficiently transfer error mitigation knowledge to circuits sharing the same ansatz but with different parameters, without requiring further training.

In addition to the Markovian noise model, we also investigate the impact of Non-Markovian noise, which, despite its relevance in real-world quantum experiments~\cite{White_2020, agarwal2023modelling, Groszkowski_2023}, has received limited attention in previous error mitigation studies.   Specifically, we consider the multi-qubit spin-boson model~\cite{10.1093/acprof:oso/9780199213900.001.0001} for phase damping to exemplify this scenario, in which a quantum system interacts with environment, namely, a heat bath, and evolves jointly. This is a potential noise happening in superconducting quantum circuits~\cite{Magazz__2018}. In this setup, depicted in Figure~\ref{fig:VQE_nm}a, the system Hamiltonian $H_S$ corresponds to the VQE circuit, while the heat bath is modeled as a bosonic system. We assume each gate in the circuit interacts independently with a bath attached locally. The bath Hamiltonian is $H_B = \sum_k \omega_k b_k^\dag b_k$. Here $b_k$ is the annihilation operator for mode $k$, and $\omega_k$ is the corresponding energy. The interaction between the system and the bath is captured by the Hamiltonian $H_{SB} = \sum_k \sigma_z \otimes [\lambda_k b_k + \lambda_k^* b_k^{\dag}]$, where $\sigma_z$ is Pauli-Z operator, and $\lambda_k \propto 1 / \sqrt{\omega_k}$. We initiate the system and bath as a product state state $\rho_0 \otimes \rho_B$, where $\rho_0$ is the initial state of the system, namely $|0\rangle \langle 0|$. The bath $\rho_B = e^{-\beta H_B} / Z$ is a Gibbs state, with $\beta = 1 / (k_B T)$ and $Z$ being a normalization factor. The evolution of the system under noisy conditions is represented as $\rho_S(t) = {\rm Tr}_B(U(t) (\rho_0 \otimes \rho_B) U^\dag(t))$. Here, $U(t) = \exp{\left[-i \int_0^t H(\tau) d\tau \right]}$ is the unitary describing the joint evolution of the whole system, with $H = H_S + H_B + H_{SB}$. Importantly, it should be noted that this noise is gate-dependent, as gate parameters rely on the evolution time of the Hamiltonians, leading to varying noise effects.

To modulate this noise, we vary the Hamiltonian evolution time within the range of $[0.05, 0.3]$, while maintaining the computational impact constant.The numerical results, as depicted in Figure~\ref{fig:VQE_nm}b, consistently demonstrate the remarkable effectiveness of our model in handling Non-Markovian noise scenarios. This performance is particularly significant because Non-Markovian noise, a common occurrence in practical quantum experiments 
~\cite{White_2020, agarwal2023modelling, Groszkowski_2023}, poses a substantial challenge for error mitigation techniques. The robustness of our model in such conditions significantly enhances its practical utility and reliability in the realm of quantum computing.

To further illustrate the effectiveness of our  model in practical scenarios, we tested its performance on the real quantum computing hardware provided by OriginQ Cloud~\cite{originq}. Based on the types of gates available on this hardware, we adopted the variational ansatz shown in Figure~\ref{fig:Vqe_real}a. The circuit is composed of three layers of $U_3$ gates and two layers of controlled-$Z$ gates.
We selected 10 circuits, varying the parameter $g$ within the range of $[1.0, 2.0)$ with a stride of 0.1. The choices of $J$, $N$, $\mathcal{M}$, and $\mathcal{S}$ were kept consistent with those used in the simulation experiments above. During the Noise-Awareness phase, we construct the fiducial circuit by replacing each $U_3$ gate with two $U_3$ gates that cancel each other out under noiseless conditions, while keeping the CZ gates unchanged, as shown in Figure~\ref{fig:Vqe_real}b.

Given the difficulty of adjusting noise strength in real experiments, we consider only one noise level ($K=1$) for our model. As the results shown in Figure~\ref{fig:Vqe_real}c, our DAEM model demonstrates superior performance in most cases compared to CDR. Note that ZNE cannot be applied in this scenario directly as it requires measurement data from circuits with varying noise strengths. For comparison, we used the method of unitary folding~\cite{giurgica2020digital} to generate data with four different noise levels for ZNE. Although our model uses data from fewer noise levels than ZNE, it still achieves significant advantages over ZNE. 

\begin{figure}[hbtp]
    \centering
    \includegraphics[width=0.3\textwidth]{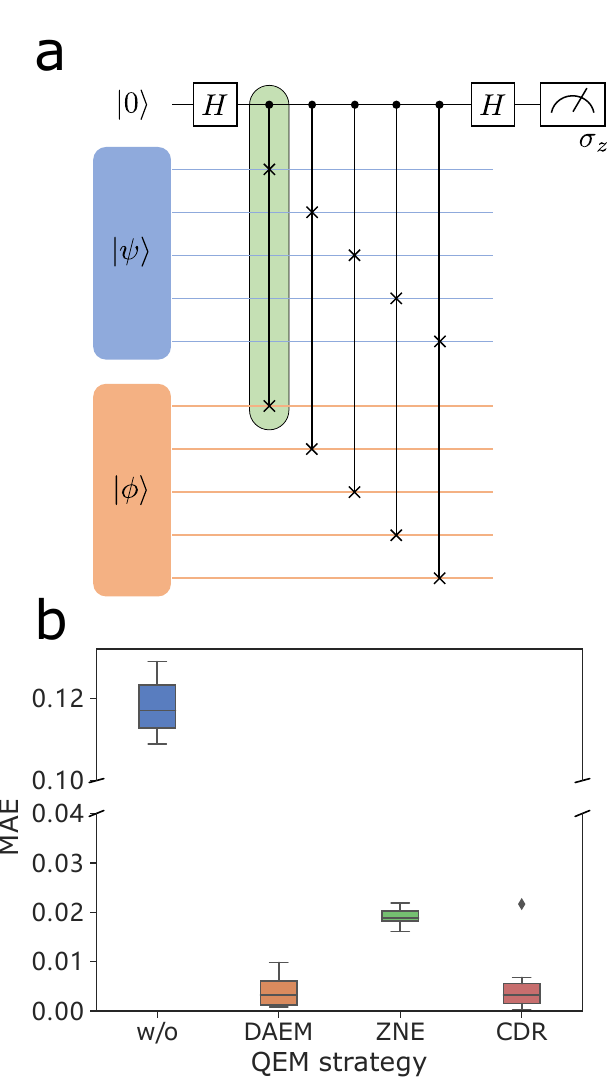}
    
    \caption{Error mitigation for the swap test. 
    a.\ The swap test circuit for comparing two 5-qubit states. The gate within the green box is the controlled-SWAP gate.
    b.\ Mean Absolute Errors (MAE) between the mitigated fidelity values and the ground truth values.
    }
    \label{fig:SwapTest}
\end{figure}



\paragraph{Swap test} The swap test is a technique used to measure the dissimilarity between two quantum states. In Figure~\ref{fig:SwapTest}a, we illustrate the circuit designed for comparing two $5$-qubit states. When executed on the quantum device, the CSWAP gate is implemented by decomposing it into three Toffoli gates, which are further decomposed using Hadamard, S, T, and CNOT gates. The details are provided in Methods section.
The circuit takes two input quantum states, $\ket{\psi}$ and $\ket{\phi}$, for comparison. It initializes the first control qubit as $\ket{0}$ and produces expectation value that equals the fidelity between two pure states, i.e., $|\langle\psi|\phi\rangle|^2$, by performing a Pauli $Z$ measurement on the first control qubit.
Here, we assume that noise takes place before each controlled-SWAP gate. Specifically, we examine phase damping channel, which can be characterized by the following equation:
\begin{equation}\label{eqn:phasedamping}
    \rho \rightarrow V_0\rho V_0^\dag + V_1\rho V_1^\dag, 
\end{equation}
with $V_0 = \begin{bmatrix} 1& 0 \\ 0 & e^{-2\lambda} \end{bmatrix}$ and $V_1 = \begin{bmatrix} 0& 0 \\ 0 & \sqrt{1 - e^{-4\lambda}} \end{bmatrix}$. $\lambda$ represents the scale of the noise and $P_i$ are three Pauli gates. We use a set of noise parameters denoted as $\{\lambda_k\}_{k=1}^K = \{0.05, 0.08, 0.12, 0.15\}$ in this experiment. 

 { In the Noise-Awareness phase, the controlled-SWAP gates in Figure~\ref{fig:SwapTest} are first decomposed into single-qubit and CNOT gates. $\mathcal{N}_\lambda(\mathcal{F})$ is constructed by replacing all single-qubit gates by two gates that cancel each other.} Specifically, for a quantum gate $G$, we replace it by $\sqrt{G^\dagger}\sqrt{G}$.  We randomly select $n$ input states $\{\sigma_{s_1}\}_{s_1=1}^n$, where each $\sigma_{s_1} = \ket{\psi}\bra{\psi}\otimes \sigma_{s_1}^1 \otimes \sigma_{s_1}^2$, with $\ket{\psi}$ being a random $1$-qubit pure state, and $\sigma_{s_1}^1$ and $\sigma_{s_1}^2$ representing two random $5$-qubit product states. 

In the Error-Mitigation phase, we evaluate our model using $20$ pairs of input states ${\rho_1,\rho_2}$ in the swap test circuit.
We collect statistics by conducting Pauli Z measurements on the first qubit within the noisy swap test circuit. These measurements are subsequently used to compute the overlap between $\rho_1$ and $\rho_2$. 
The noisy expectation values obtained from these measurements are then input into the trained neural model, which produces mitigated values as output. 
In Figure~\ref{fig:SwapTest}b, we present Mean Absolute Errors (MAE) between the mitigated values and the ground truth values, providing a comparative analysis with two other quantum error mitigation techniques, ZNE and CDR. The performance of DAEM stands out, showcasing significant improvements over ZNE and demonstrating comparable performance with CDR.

\paragraph{Quantum approximate optimization algorithms}

\begin{figure}[hbtp]
    \centering    \includegraphics[width=0.47\textwidth]{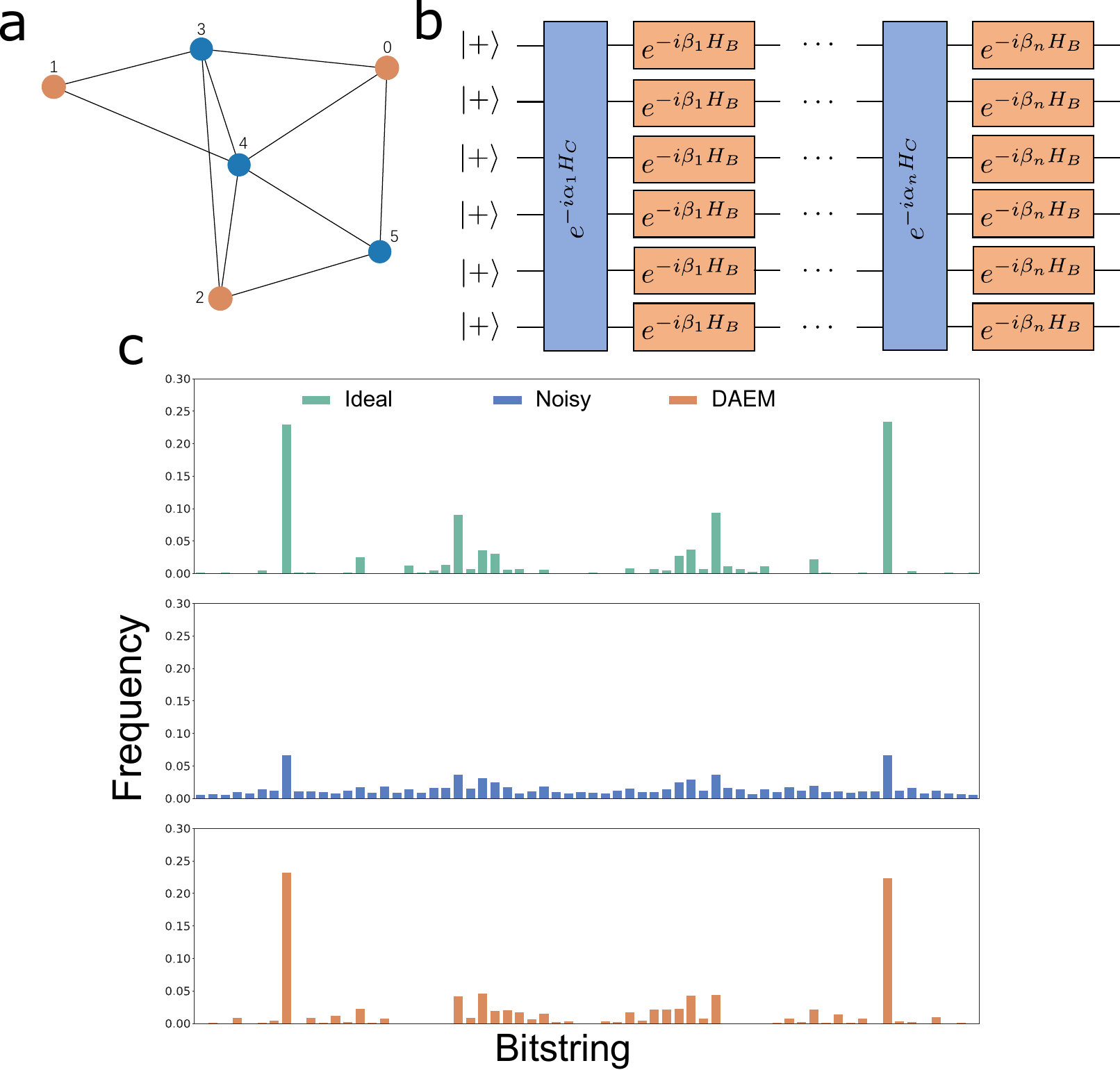}
    \caption{Error mitigation for quantum approximate optimization. 
    a.\ An instance of a graph for for the Max-cut problem.
    b.\ The variational ansatz for implementing QAOA algorithm.
    c.\ Ideal, Noisy and Mitigated frequency of measurement results.
    }
    \label{fig:qaoa}
\end{figure}
QAOA~\cite{zhou2020quantum} is a quantum algorithm specifically designed for solving combinatorial optimization problems. The core of this algorithm involves encoding the objective function of the target optimization problem into a Hamiltonian, and trains an elaborately designed parameterized circuit to approximate the ground state. The final solution is derived by sampling bitstrings from the circuit's output in the computational basis. However, when running a well-trained QAOA circuit on noisy quantum computers, the resulting output distribution deviates from the ideal scenario, which results in less accurate solutions. Hence, our goal is to mitigate this noise-induced bias in the output distribution, thereby providing experimenters with more precise solutions.

In this specific application, we focus on implementing QAOA for the maximum cut (Max-cut) problem~\cite{guerreschi2019qaoa}. The goal is to find a bi-partition of a graph $G$, namely subsets $A$ and $B$, in which the partition contains the maximum number of edges. This can be defined as an optimization with objective
\begin{equation}
    \max_{\boldsymbol z} L(\boldsymbol{z}) = \frac{1}{2}\sum_{(i, j)\in E} (1 - z_i z_j),
\end{equation}
where $i$, $j$ denote the indices of vertices, $(i, j)$ represents the edge connecting vertex $i$ and vertex $j$, and $E$ is the set containing all edges of the graph. If vertex $i$ belongs to subset $A$, then $z_j = 1$, otherwise $z_j = 0$. We provide an instance of $G$ with 6 vertices in Figure.~\ref{fig:qaoa}a. 
The corresponding Hamiltonian of this problem in QAOA can be described by the following:
\begin{equation}
    H_C = \frac{1}{2} \sum_{(i, j)\in E} (I - Z_i Z_j),
\end{equation}
where $Z$ is Pauli-Z operator. The circuit for QAOA, shown in Figure~\ref{fig:qaoa}b, typically comprises two sets of parameterized quantum gates, alternating between a mixing operator $H_B = \sum_{n=1}^N \sigma^x_n$  and the problem-specific cost operator $H_C$.   In the Noise-awareness phase, to generate fiducial process,  after replacing all single-qubit gates with identity gates, the CNOT gates automatically cancel each other. In this case, the fiducial process is trivially identity ideally, i.e., $\mathcal N_0(\mathcal F) = I$. Again, we execute the fiducial process on noisy quantum devices with different noise parameters to acquire noisy bitstring distributions. Besides, we sample input states $|\psi\rangle$ in computational basis to obtain labels for training.
In Figure~\ref{fig:qaoa}c, we present the mitigated results concerning the output state of a trained QAOA circuit applied to the graph depicted in Figure~\ref{fig:qaoa}a. Here, we consider the depolarizing noise model and it can be described by 
\begin{equation}\label{eqn:amplitudedamping}
    \rho \rightarrow (1-\lambda)\rho + \frac{\lambda}{4^N-1}\sum_i P_i\rho P_i, 
\end{equation}
where $P_i$ are the $4^N-1$ Pauli gates excluding the identity gate. It's evident that the mitigated frequency of measurement results closely approximates the ideal scenario, signifying that we can obtain more reliable solutions to the original Max-Cut problem through our DAEM model. It's worth noting that both ZNE and CDR are designed specifically for mitigating errors in expectation values, and they cannot be applied directly to the probability distribution of measurement results, in contrast to our proposed DAEM.

\begin{figure}
    \centering
    \includegraphics[width=0.35\textwidth]{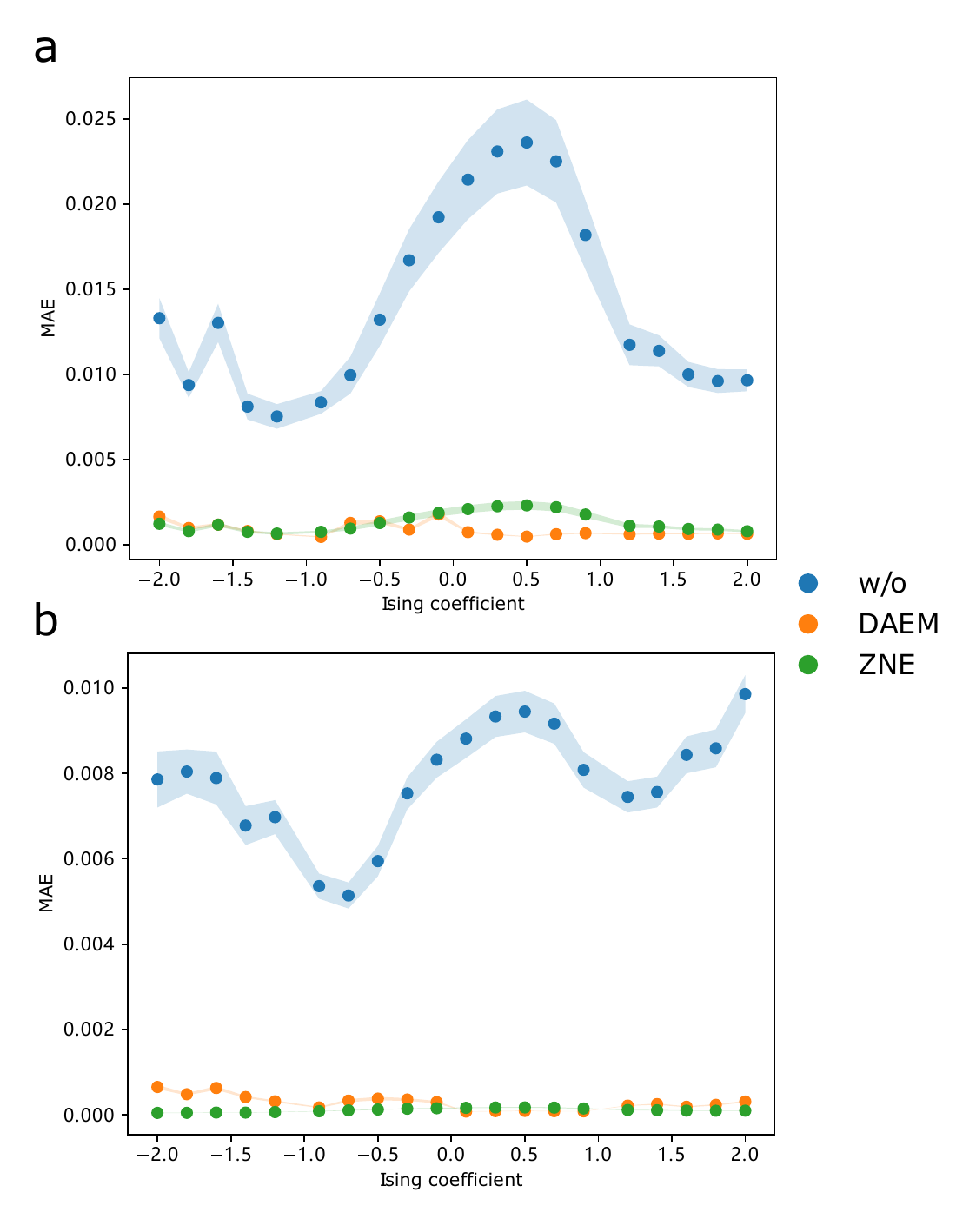}
   
    \caption{Error mitigation for a quantum spin dynamics. 
    a.\  Mean Absolute Errors (MAE) between the mitigated measurement expectation values for phase damping noise model and ideal expectation values.
    b.\ Mean Absolute Errors (MAE) between the mitigated measurement expectation values for amplitude damping noise model and ideal expectation values.
    }
    \label{fig:dynamics}
\end{figure}

\subsection{Error mitigation for many-body dynamics} 
Our model works for quantum processes beyond the circuit model. It applies to, for example, the dynamics of physical systems. In this section, we delve into the challenge of error mitigation within the domain of many-body dynamics, which is fundamental to various applications in quantum physics and materials science.

Here, our focus is on the dynamics of a 50-qubit quantum system with an Ising Hamiltonian $H_{\text{Ising}}$, descirbed in Eqn.~\ref{eqn:Ising}. We consider the whole system's evolution for time $t$, given as $U=\exp(-\textit{i} H_{\text{Ising}} t)$. This specific process is characterized by the following parameters: $J=1$, $g=2$, and a time duration of $t=5$. For the initial states involved in this Ising Hamiltonian evolution, we have selected the ground states of the Ising model, varying $J$ within the range $[-2, 2]$, while keeping $g$ constant at $1$. 
To simulate these processes, we employ a combination of two powerful techniques: the density matrix renormalization group (DMRG)~\cite{white1992density} and time-evolving block decimation (TEBD)~\cite{white2004real, daley2004time}. In this setup, noise is introduced after the completion of the unitary quantum process. Specifically, we evaluate our model's performance under two distinct noise models: phase damping and amplitude damping channels.  For the set of measurements $\mathcal{M}$, we also consider all two-qubit Pauli measurements on nearest-neighbor qubits. The results, corresponding to different values of $g$ in the input states, are presented in Figure~\ref{fig:dynamics}. We can observe that both our model and ZNE have achieved nearly perfect mitigation, as evidenced by the  MAE between the mitigated expectation values and the ground-truth values, which are close to zero. We conjecture the reason for the nearly perfect performance of ZNE in this experiment is that no SPAM errors have been introduced. This makes the actual measurement expectation values decay quadratically with respect to the noise parameters, perfectly fitting the ansatz of ZNE.  It is important to highlight that ZNE relies on precise knowledge of the noise parameters corresponding to the noisy measurement data, whereas our DAEM model does not have this requirement. CDR is a technique tailored for quantum circuits, and therefore, it cannot be employed to mitigate errors in spin-system dynamics.

\begin{figure*}
    \centering
    \includegraphics[width=0.6\textwidth]{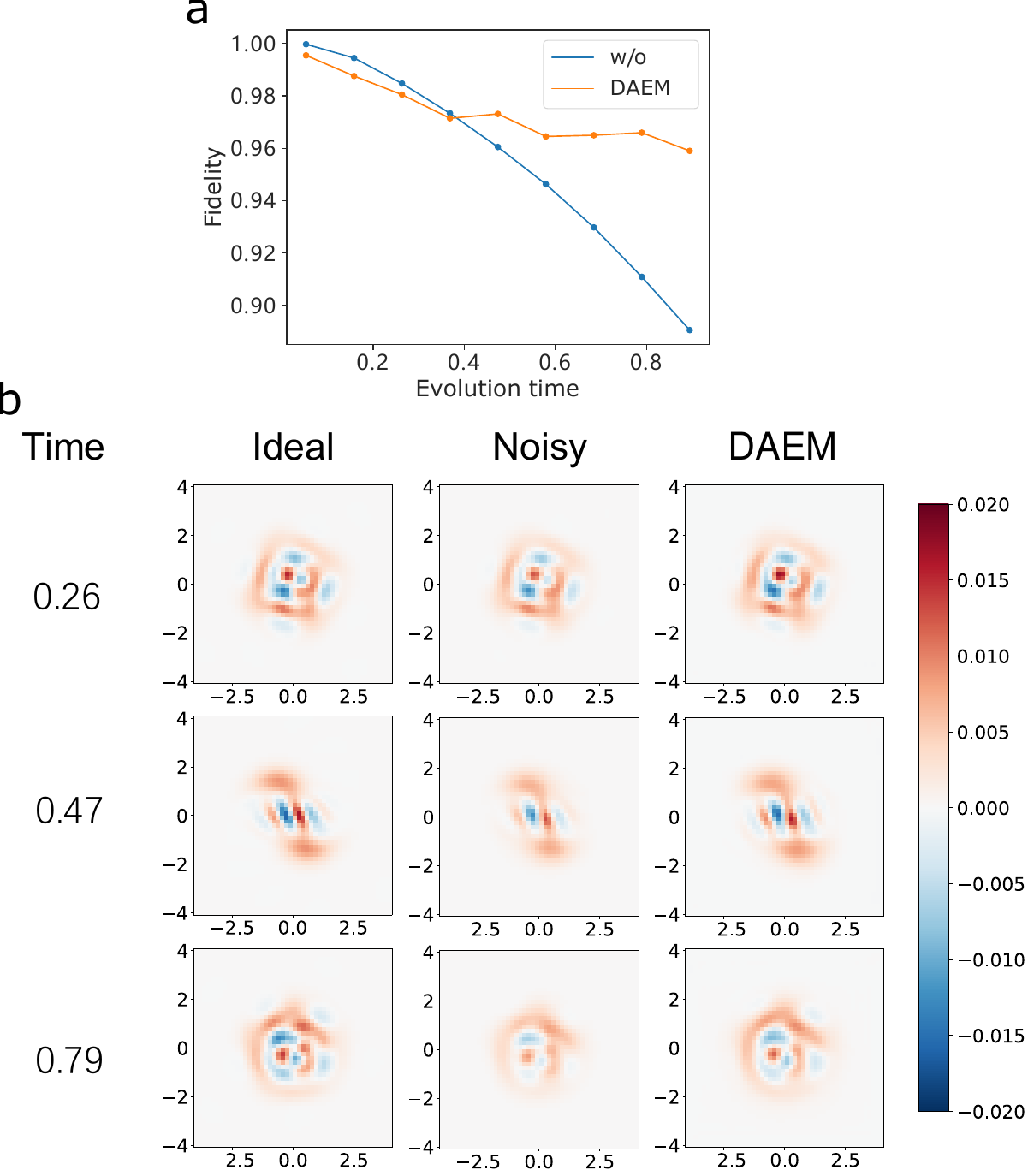}
    \caption{Error mitigation for the Kerr gate in a continuous variable system. 
    a.\ Fidelity values between the noisy/mitigated state and the ideal output state.
    b.\ Snapshots of the point-wise measurement results of the state at different time points.
    }
    \label{fig:continuous_variable}
\end{figure*}

\subsection{Error mitigation for continuous-variable processes}\label{sec:cv} Continuous-variable quantum systems have demonstrated their potential in diverse applications including quantum cryptography~\cite{jouguet2013experimental} and quantum computing~\cite{lloyd1999quantum}. However, despite the growing significance of this type of systems, no prior research has discussed the issue of error mitigation within continuous-variable quantum systems as far as we know. In this section, we applied our proposed DAEM model to address this long-unexplored challenge first.

We assess the effectiveness of our method on the dynamics induced by Kerr's nonlinear interaction~\cite{dykman2012} 
, which is important for continuous variable quantum computing~\cite{lloyd1999quantum}. Consider a quantum system initially prepared in a coherent state $\ket{\alpha}$ and subjected to the Kerr Hamiltonian $H_{\rm kerr}=\pi\hat{a}^{\dagger\, 2}\hat{a}^2$, where $\hat{a}$ and $\hat{a}^\dagger$ represent the annihilation and creation operators. In this scenario, we model the noisy process by a lossy open system, whose dynamics is described by the Lindblad master equation:
\begin{equation}
    \dot{\rho}=-\text{i}[H_{\rm kerr},\rho]+\lambda \mathcal D(\hat{a})(\rho).
\end{equation}
Here, $\mathcal D(\hat{a})(\rho) := \hat{a}\rho\hat{a}^\dagger-\frac{1}{2}(\rho \hat{a}^\dagger\hat{a}+\hat{a}^\dagger\hat{a}\rho)$, and $\lambda$ represents the loss rate. Our objective is to mitigate the errors in this process, making it  closely resemble the ideal closed-system dynamics governed by Schrödinger's equation with Hamiltonian $H_{\rm kerr}$. In this setting, we consider the measurement results associated with the point-wise Wigner function~\cite{PhysRevLett.78.2547,PhysRevLett.89.200402}. 
In our numerical experiments, we initialize the state as  coherent state $\ket{\alpha}$ with $\alpha = 1.5$ and dimension $N = 15$. We vary the evolution time over the interval $t \in [0,1]$, considering different loss rates $\lambda \in \{0.6,0.65,0.7,0.75,0.8\}$. 
To train the neural model within DAEM, we construct the fiducial process $\mathcal{N}_\lambda(\mathcal{F})$ by implementing the evolution of two inverse Hamiltonians, ensuring that the overall effect on the input state is an identity operation under noiseless conditions. Specifically, assuming a total evolution time of \( t_0 \), the state evolves with the Hamiltonian \( H_{\text{kerr}} \) for \( t \leq t_0 / 2 \), and with \( -H_{\text{kerr}} \) for \( t_0 / 2 \leq t \leq t_0 \). We assess the effectiveness of our DAEM by computing the fidelities between the mitigated states and their noiseless counterparts, employing the values of the point-wise Wigner function. As shown in the numerical results presented in Figure~\ref{fig:continuous_variable}a, the fidelity between the state affected by the noise and the ideal state decreases rapidly as time increases. In contrast, our method excels in mitigating this effect, resulting in a dramatic improvement in fidelity. We also present snapshots of the state at different time points in Figure~\ref{fig:continuous_variable}b, and our mitigated point-wise measurement results are notably closer to the ideal ones, particularly for longer evolution.

\section{Discussion}

The workhorse of our model is the quantum data augmentation method, which generates the training data by letting the unknown noise act on a set of ideal fiducial processes. This  technique is not only applicable to quantum error mitigation, but also  to other tasks in quantum information processing, including in particular the task of enhancing parameter estimation in quantum metrology~\cite{giovannetti2006quantum}. By combining quantum data augmentation with the representational capability of deep neural networks, our model becomes able to effectively handle complex noise scenarios. For example, it deals effectively with Non-Markovian noise ({\em viz.} Figure~\ref{fig:VQE}e), 
for which other neural models, like  ZNE, tend to perform poorly due to reliance on a predefined extrapolation algorithm.  

Our model  also offers  appealing features compared to conventional error mitigation methods. In CDR, an error mitigation model is better trained with classically simulated quantum circuits that resemble the target circuit \cite{czarnik2021error}.
 Therefore, the effectiveness of CDR can depend heavily on how closely the training circuits match the target circuit, and achieving such a close match might not always be feasible in practical experiments. In contrast, our proposed DAEM model conducts training directly on the data collected from the hardware, targeting the specific noise to be mitigated. This approach ensures more accurate and effective error mitigation tailored to the actual noise characteristics of the hardware. It is also worth observing that CDR  and ZNE are effective at mitigating  expectation values, but generally less effective at mitigating the whole probability distribution of the measurement outcomes,  a task that is necessary in quantum algorithms like QAOA.  For noise preservation, our approach shares a similar spirit with Ref.~\cite{sack2024large}, which targets for mitigating errors in QAOA circuits. {
 Ref.~\cite{sack2024large} works by considering a modified version of the original circuit, where all single-qubit $R_Z$ gates in the cost gates are ignored. This corresponds to a modified problem Hamiltonian whose ZZ coupling strength is zero, and thus the modified circuit can be simulated efficiently with a classical computer.
 The modified circuit is then executed on a (noisy) quantum computer. As the (pairwise) CNOT gates are left unchanged in the modified circuits, the outputs of the real-device execution can be compared with the simulation to learn the pattern of noise propagation.} Likewise in our framework, we elaborately design our data augmentation strategy that preserves the skeletons of the circuits. Our approach can also be compared to probabilistic error cancellation~\cite{temme2017error, endo2018practical,van2023probabilistic}, which estimates noise-free expectation values by representing them as linear combinations of expectation values from a set of noisy quantum circuits.  
  To work out the appropriate decomposition, this method requires a  tomography of the noise, 
  thus resulting in an overhead in  sample complexity.  A benefit of our approach is that it removes the need of tomography and replaces it with the quantum data augmentation procedure, which is generally less demanding in terms of number of measurement settings. 

In terms of scalability, our model can potentially be scaled to larger systems. We conduct experiments on mitigating errors of circuits with different number of qubits and different circuit depth, with fixed number of training data. Results show that our model can keep stable performance with respect to different circuit configurations. Further information can be found in Supplementary Note 4.

Finally, our model has the potential for extension to mitigate a broader range of realistic quantum errors, including crosstalk errors~\cite{9193969, Sarovar2020detectingcrosstalk}, which are common in  quantum computing systems.  Crosstalk errors result from hardware imperfections that violate the assumption of locality and independence of quantum operations, and are therefore challenging to model~\cite{10.1145/3373376.3378477}.  
  Despite these challenges, error mitigation for crosstalk errors could  become  approachable in our framework, which does not require prior error modeling.

\section{Methods}
\subsection{Neural Model in DAEM}
Our error mitigation is model-agnostic thus the structure of the neural model can be flexibly chosen, ranging from simple linear models~\cite{hastie01statisticallearning}, multi-layer perceptrons (MLP)~\cite{haykin1994neural}, to deep neural networks like convolutional neural networks~\cite{he2015deep} and Transformers~\cite{10.5555/3295222.3295349}. In practice, we adopt a problem-aware strategy to design the specific construction of the model. In general, we prefer non-linear models for they have stronger ability to capture the intrinsic characteristics of various noise models.

For error mitigation in quantum algorithms and many-body dynamics, we use MLP as the architecture of the neural model. The neural network is composed of multiple layers of fully connected neurons. Each neuron involves one linear transform followed by one non-linear activation. The stack of neurons allows for complex non-linear function fitting, which is powerful for estimating the expectation values and probability distributions in our error mitigation settings. The model's inputs are parameters indicating the target circuit for mitigation, the observable to be measured, and the measurement statistics. The model's output is either a real number or a probability distribution obtained by passing through a softmax function, depending on the measurement statistics to be mitigated. The cost function for mitigating expectation values is $L_2$ loss, namely
\begin{equation}
    \mathcal L (\boldsymbol y, \boldsymbol y_{\text{fid}}) = \frac{1}{n}\sum_{i = 1}^n (y^{(i)} - y_{\text{fid}}^{(i)})^2,
\end{equation}
where $\boldsymbol{y}$ denotes the output of the neural model and $\boldsymbol{y}_{\text{fid}}$ are the observable expectation values generated from data augmentation using the fiducial channel. The cost function for mitigating probability distribution is the average relative entropy~\cite{10.5555/1146355}, defined as
\begin{equation}
    \mathcal L(\boldsymbol{p}, \boldsymbol{p}_{\text{fid}}) = \frac{1}{n} \sum_{i = 1}^n \sum_x p^{(i)}(x) \log \left(\frac{p^{(i)}(x)}{p^{(i)}_{\text{fid}}(x)}\right),
\end{equation}
where $\boldsymbol{p}$ are probability distribution predicted by our model, and $\boldsymbol{p}_{\text{fid}}$ are distribution obtained from fiducial channel by sampling 10000 shots. 

To mitigate errors in continuous-variable processes, we adopt U-Net~\cite{ronneberger2015unet}, a convolutional neural network originally designed for image segmentation~\cite{9356353}, to be the neural model denoising the 2-dimensional Wigner function. U-Net possesses the strong ability to extract spatial features and construct 2-dimensional distributions. In this sense, it helps learn the distribution of the point-wise Wigner function. The inputs to the model are evolution time and Wigner functions corresponding to different photon loss rates. The output is a single 2-dimensional feature map, which represents the denoised Wigner function. To train the neural model, we use $L_1$ loss as cost function, defined as
\begin{equation}
    \mathcal L (\boldsymbol y, \boldsymbol y_{\text{fid}}) = \frac{1}{n}\sum_{i = 1}^n \left|y^{(i)} - y_{\text{fid}}^{(i)}\right|.
\end{equation}
It encourages sparse output distribution, which conforms to the Wigner quasiprobability distribution of our target states. More details can be found in Supplementary Note 1.

\subsection{Data augmentation strategy}

In practice, the noise-free labels are not available unless we know the exact noiseless output states of the circuits. However, if the input and output states are the same, or under known transformation in the noise-free scenario, we can directly measure the input states to generate labels. Here, we introduce a fiducial process, i.e., $\mathcal N_{\lambda} (\mathcal F)$, to achieve this goal. { The process is trivially identity or contains only CNOT gates that can be absorbed into observables in the noise-free conditions, but shares similar noise pattern as the target process in a noisy quantum device.} To generate the training set for Noise-Awareness phase, we send input states through the fiducial process, measure the noisy outputs as data and measure the original input states as labels. Note that the input states can be either noisy or noiseless.

We detail the choice of input states and specification of $\mathcal N_{\lambda} (\mathcal F)$ for different applications as follows.

\begin{figure}
    \centering
    \includegraphics[width=0.9\linewidth]{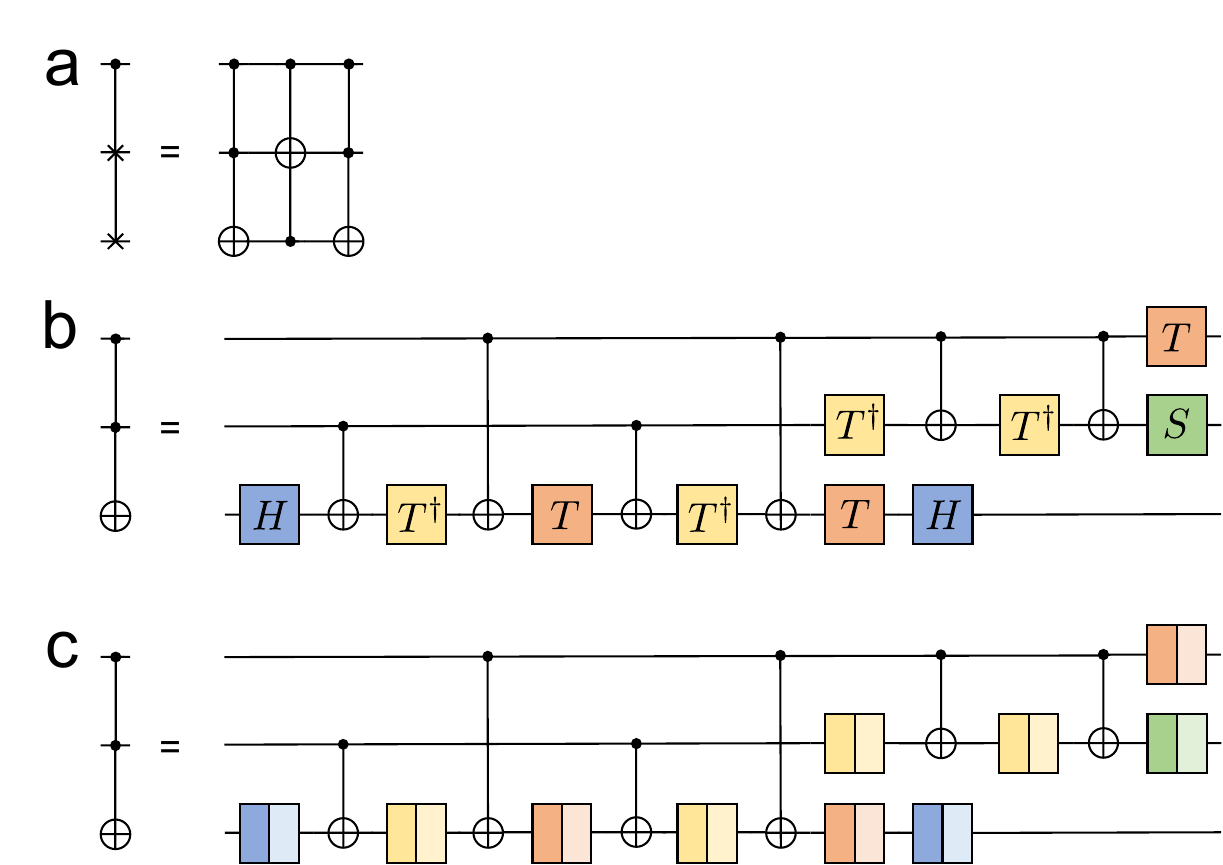}
    \caption{CSWAP gate decomposition. a.\ CSWAP gate is decomposed into 3 Toffoli gates. b.\ Further decomposition of Toffoli gate into single-qubit and CNOT gates. c.\ The corresponding fiducial circuit for training.}
    \label{fig:cswap}
\end{figure}

For swap test circuits, the states to be compared are pure. 
We decompose the CSWAP gates into 3 Toffoli gates. The Toffoli gates are further transformed into CNOT and single qubit gates, as shown in Figure \ref{fig:cswap}. 

In Noise-Awarenes phase, we sample 170 random pairs of pure states from the single-qubit Haar measure, in which 100 are used for training, 50 for validation, and 20 for Error-Mitigation phase. For ancilla, we choose random single-qubit mixed states.  { We construct the fiducial process by replacing every single-qubit gate $G$ with identity gate, which is implemented by $\sqrt{G^{\dagger}} \sqrt{G}$, leaving the CNOT gates unchanged.} This results in a channel $\mathcal N_0(\mathcal F) = U$, in which $U$ describes the effects of all CNOT gates in the original circuit. Then we execute the fiducial process in noisy environment with varying noise parameters, and measure the noisy outputs using observable $M = Z_1$, which denotes the Pauli-Z observable on the ancilla qubit, and calculate the expectation values. Meanwhile, we measure the input states with observable $\tilde M = U^\dagger Z_1 U$. The measurement expectation values are the corresponding labels.

In VQEs, the augmentation strategy is generally the same as in Swap test. One difference is that, rather than pure states, we randomly sample 100 states as inputs in the Noise-Awareness phase. Whereas in the Error-Mitigation phase, the input states are chosen to be ground state $|0\rangle$.

For QAOA circuits, note that the distributions of output bitstrings possess symmetry, e.g., if 00011 is one solution, 11100 should also be a solution. To boost the performance of the neural model, we want to make the output distributions of the dataset in Noise-Awareness phase more aligned with those in the Error-Mitigation phase, i.e., the output distributions in the training set also possess this symmetry. It can be mathematically described as $X^{\otimes n}|\psi\rangle = |\psi\rangle$, where $X$ is Pauli-X operator. This shows that the input states $|\psi\rangle$ are the eigenvectors of $X^{\otimes n}$ with eigenvalue 1. In our implementation, we sample 100 vectors from the eigenspace of $X^{\otimes n}$ with eigenvalue 1 as the input states. In the Noise-awareness phase, to generate fiducial process,  after replacing all single-qubit gates with identity gates, the CNOT gates automatically cancel each other. In this case, the fiducial process is trivially identity ideally, i.e., $\mathcal N_0(\mathcal F) = I$. Again, we execute the fiducial process on noisy quantum devices with different noise parameters to acquire noisy bitstring distributions. Besides, we sample input states $|\psi\rangle$ in computational basis to obtain labels for training.

For spin systems, the input states in both Noise-Awareness phase and Error-Mitigation phase are sampled from the same distribution. We have 100 different states for training in the first phase and 20 for testing in the second phase. The fiducial process is constructed by simply setting $H_{\text{Ising}} = I$.

For continuous-variable processes, to generate initial states, we first record intermediate states during noisy evolution of $H_{\text{kerr}}$ in the time interval $t\in [0, 1]$, each with timestep 0.05. With this procedure, we obtain 20 noisy states. Next, for each state, we evolve it with fiducial process $\mathcal N_0(\mathcal F) = I$ under different loss rates. { The fiducial process is generated by $H_{\text{kerr}}$ followed by $-H_{\text{kerr}}$, which can be simulated on hardware by~\cite{PhysRevLett.131.110603}. The evolution times of the states are uniformly chosen in the range $[0, 1]$ with a gap of 0.1.}

\section{DATA AVAILABILITY}

Data sets utilized during this study are available from the corresponding author upon reasonable request. 

\section{Code AVAILABILITY}

The code supporting the findings of this study is available from the corresponding author upon reasonable request.

\section{COMPETING INTERESTS}
The authors declare no competing financial or non-financial  interests.

\section{author contributions}
M.\ L., Y.\ Z.\ and  Y.\ Y.\ established the key idea in the paper. M.\ L.\ developed the neural network model and did the numerical experiments. Y.\ Z.\ wrote the draft paper. All the authors contributed to the preparation of the paper.

{\em Acknowledgement.} We thank Qiushi Liu, Ya-Dong Wu for the helpful discussions.
This work is supported by the Guangdong Basic and Applied Basic Research Foundation (Project No. 2022A1515010340), Guangdong Provincial Quantum Science Strategic Initiative (No. GDZX2303007), and by the Hong Kong Research Grant Council (RGC) through the Early Career Scheme (ECS) grant 27310822, the General Research Fund (GRF) grants no.\ 17303923, no.\ 17300918 and no.\ 17307520, and the Senior Research Fellowship Scheme SRFS2021-7S02, and the John Templeton Foundation through grant  62312, the Quantum Information Structure of Spacetime (qiss.fr). Research at the Perimeter Institute is supported by the Government of Canada through the Department of Innovation, Science and Economic Development Canada and by the Province of Ontario through the Ministry of Research, Innovation and Science. The opinions expressed in this publication are those of the authors and do not necessarily reflect the views of the John Templeton Foundation.

{\em Note added.} After the completion of this work, another work~\cite{danila2023echo} appeared where  a data generation technique is employed to mitigate errors in the transverse field Ising model. This approach can be viewed as a specific means of obtaining the fiducial process.

\bibliography{sn-bibliography}

\appendix
\onecolumngrid

\section{Implementation details of DAEM}
\subsection{Dataset construction}
The algorithm for constructing dataset both in the Noise-Awareness phase and in Error-Mitigation phase is shown in Algorithm \ref{algorithm:data}.

\begin{algorithm}
\caption{Dataset construction for DAEM}\label{algorithm:data}
\KwData{indicator of the phase in DAEM $p$, the set of input states $\mathcal S$, observables $\mathcal M$, the target quantum processes $\mathcal N_\lambda(\mathcal E)$, property of the target process $g$, noise parameters $\lambda$, { indicator of the} measurement statistics $m$.}
\KwResult{dataset $\mathcal D$.}
Initialize $\mathcal D \leftarrow \emptyset$\;
\eIf{$p = \text{Noise-Awareness phase}$}{
Construct fiducial process $\mathcal N_\lambda(\mathcal F)$\;
$\mathcal N_\lambda \leftarrow \mathcal N_\lambda(\mathcal F)$\;
}{
$\mathcal N_\lambda \leftarrow \mathcal N_\lambda(\mathcal E)$\;
}
\For{each $\rho_s\in \mathcal S$}{
\For{each $\boldsymbol M \in \mathcal M$}{
\tcp{Construct data.}
$\boldsymbol{P} \leftarrow \emptyset$\;
\For{each $\lambda_k \in \lambda$}{
Pass the state $\rho_s$ into the quantum process $\mathcal N_{\lambda_k}$ and obtain the output state $\rho_{o} = \mathcal N_{\lambda_k}(\rho_s)$\;
\eIf{$m = \text{expectation value}$}{
Measure the output state $\rho_o$ with observable $\boldsymbol{M}$ and calculate the expectation value $\boldsymbol{p} = \tr(\boldsymbol{M}\rho_o)$\;
}{
Measure the output state $\rho_o$ with observable $\boldsymbol{M}$ obtain the output distribution of the bitstrings $\boldsymbol{p}$\;
}
$\boldsymbol{P} \leftarrow \boldsymbol{P} \cup \{\boldsymbol{p}\}$\;
}
\tcp{Construct labels.}
\eIf{$p = \text{Noise-Awareness phase}$}{
\eIf{$\mathcal N_0(\mathcal F) \neq I$}{
$U \leftarrow \mathcal N_0(\mathcal F)$\;
$M_0 \leftarrow U^\dagger \boldsymbol{M} U$\;
}{
$M_0 \leftarrow \boldsymbol{M}$\;
}
Measure input state $\rho_s$ with observable $M_0$ and obtain the measurement statistics $\boldsymbol{p}_0$\;
}{
Simulate the ideal output state for evaluation $\rho_i = \mathcal N_0(\mathcal E)$\;
Measure output state $\rho_i$ with observable $\boldsymbol{M}$ and obtain the measurement statistics $\boldsymbol{p}_0$\;
}
$\mathcal D \leftarrow \mathcal D \cup \{(g, \boldsymbol{M}, \boldsymbol{P}, \boldsymbol{p}_0)\}$\;
}
}

\end{algorithm}

In the Noise-Awareness phase, we generate input states $\mathcal S_{\text{NA}} \subset \mathcal S$ and fiducial process $\mathcal N_\lambda (\mathcal F)$ as introduced in the main text. We choose 100 states for training and 50 for validation. The state $\rho_s$ is passed into the fiducial process with various noise parameters $\{\lambda_k\}$ described in the main text. For each output state with noise parameter $\lambda_k$, we measure every adjacent qubits using all two-local Pauli obervables $\boldsymbol{M}_i \in \mathcal M$, and obtain the measurement statistics $\boldsymbol{p}^{k}_{is}$. Note that in the continuous variable experiment, the output is a Wigner quasi-probability distribution, which requires no observable for measurement, thus we set $\mathcal M = \{I\}$. The measurement statistics with different noise parameters are collected as a vector $\boldsymbol{P}_{is} = \left(\boldsymbol{p}^{(1)}_{is},\dots,\boldsymbol{p}^{(K)}_{is}\right)$. Meanwhile, we note down the property of the target process as $g$. For VQE and spin system experiments, $g$ stands for the coefficient of the Hamiltonian. For QAOA and swap test experiments, $g$ is purely integer indicating individual process. For continuous variable process, $g$ is the evolution time of the state. Besides, to generate labels, we measure the input state $\rho_s$ and obtain the measurement statistics $\boldsymbol{p}^{(0)}_{is}$. After recording these data, we put them together to form the dataset $\mathcal D_{\text{NA}} = \left\{g_i, \boldsymbol{M}_i, \boldsymbol{P}_{is}, \boldsymbol{p}^{(0)}_{is}\right\}$.

In Error-Mitigation phase, the selection of input states $\mathcal S_{\text{EM}} \subset \mathcal S$ and quantum process $\mathcal N_\lambda(\mathcal E)$ conforms to the experiment settings. The state is passes through the quantum processes with the same noise parameters as in the Noise-Awareness phase, and the output is measured using the same observables. The dataset is formed as $\mathcal D_{\text{EM}} = \left\{g_i, \boldsymbol{M}_i, \boldsymbol{P}_{is}, \boldsymbol{p}^{(0)}_{is}\right\}$. Whereas here $\boldsymbol{p}^{(0)}_{is}$ is obtained by classical simulation, which is used only to evaluate the performance.

\subsection{Neural model in DAEM}
The neural model in DAEM is implemented using PyTorch~\cite{paszke2019pytorch} package. The specific structure of the neural model is determined by input data format. For mitigating errors in quanum algorithms and spin-system dynamics, we use multi-layer perceptron (MLP). For mitigating errors in continuous variable processes, we use a modified U-Net.

\begin{figure}
    \centering
    \includegraphics[width=0.6\textwidth]{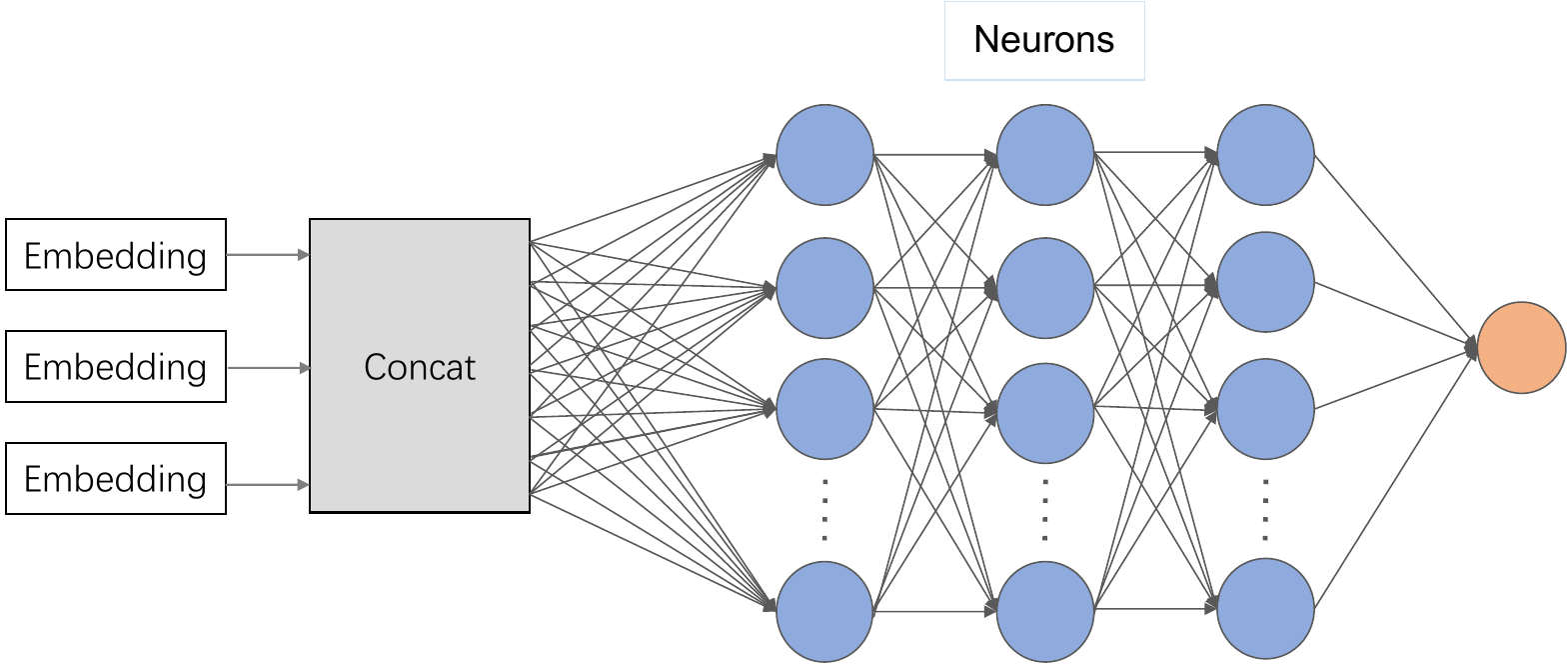}
   
    \caption{Structure of MLP used for mitigating errors in quantum algorithms and Spin-system dynamics.
    }
    \label{fig:mlp}
\end{figure}

\paragraph{MLP}
The MLP for mitigating errors in quantum algorithms and spin system dynamics consists of 1 embedding block and 4 layers of neurons. The dimensions of hidden layers are 512, 1024, 1024 respectively. Besides, the nonlinear activation in our implementation is Mish~\cite{misra2020mish}, an advanced activation function widely applied in modern neural network designs. The structure of the network is shown in Figure \ref{fig:mlp}. 
The input of the network is $(g_i, \boldsymbol M_i, \boldsymbol{P}_i)$. The embedding block maps the values of $g_i$, $\boldsymbol{P}_i$ into 128 dimensional vectors. For $M_i$, we separate the real and image part, concatenate them together and flatten the whole matrix into a vector. The vector is then transformed by the embedding block into a 128 dimensional vector. After embedding, the subsequent neurons takes the embedded vectors as input and outputs the mitigated measurements. The output of the network is the mitigated measurement statistics $\hat{\boldsymbol{p}}^{(0)}_i$.

For training, we use Adam~\cite{kingma2017adam} as optimizer with initial learning rate 2e-4 and train for 300 epochs. The batch size is chosen to be 64. The training procedure follows the algorithm described in Algorithm \ref{algorithm:train}.

\begin{algorithm}
\caption{Training of neural model in Noise-Awareness phase.}\label{algorithm:train}
\KwData{dataset $\mathcal D_{\text{NA}}$, number of epochs $E$, batch size $B$, neural model $f_{\boldsymbol{\theta}}$.}
\KwResult{Neural model $f_{\boldsymbol{\theta}}$ with trained parameters $\boldsymbol{\theta}$.}
Initialize random parameters $\boldsymbol{\theta}$, $itr = 0$\;
\While{$itr < E$}{
$\mathcal D \leftarrow \mathcal{D_{\text{NA}}}$\;
\While{$\mathcal D \neq \emptyset$}{
Randomly sample $B$ unrepeated samples from the dataset $\boldsymbol{d} \sim \mathcal D_{\text{NA}}$\;
Separate the data $\boldsymbol{x} = \{g_i, \boldsymbol{M}_i, \boldsymbol{P}_i\}_{i = 1}^B$ and labels $\boldsymbol{y} = \{\boldsymbol{p}^{(0)}_i\}_{i=1}^B$\;
Input data to the neural model and obtain predicted results $\hat{\boldsymbol{p}}^{(0)} = f_{\boldsymbol{\theta}}(\boldsymbol{x})$\;
$\mathcal D \leftarrow \mathcal D \setminus {\boldsymbol{d}}$\;
Calculate the loss using the cost function according to the mitigation task as $l = \mathcal L(\hat{\boldsymbol{p}}^{(0)}, \boldsymbol{y})$\;
Update the parameters of the model using Adam optimizer $\boldsymbol{\theta} \leftarrow \boldsymbol{\theta} - \text{Adam} \left(\nabla_{\boldsymbol{\theta}} l \right)$\;
}
$itr \leftarrow itr + 1$\;
}
\end{algorithm}

\begin{figure}
    \centering
    \includegraphics[width=0.7\textwidth]{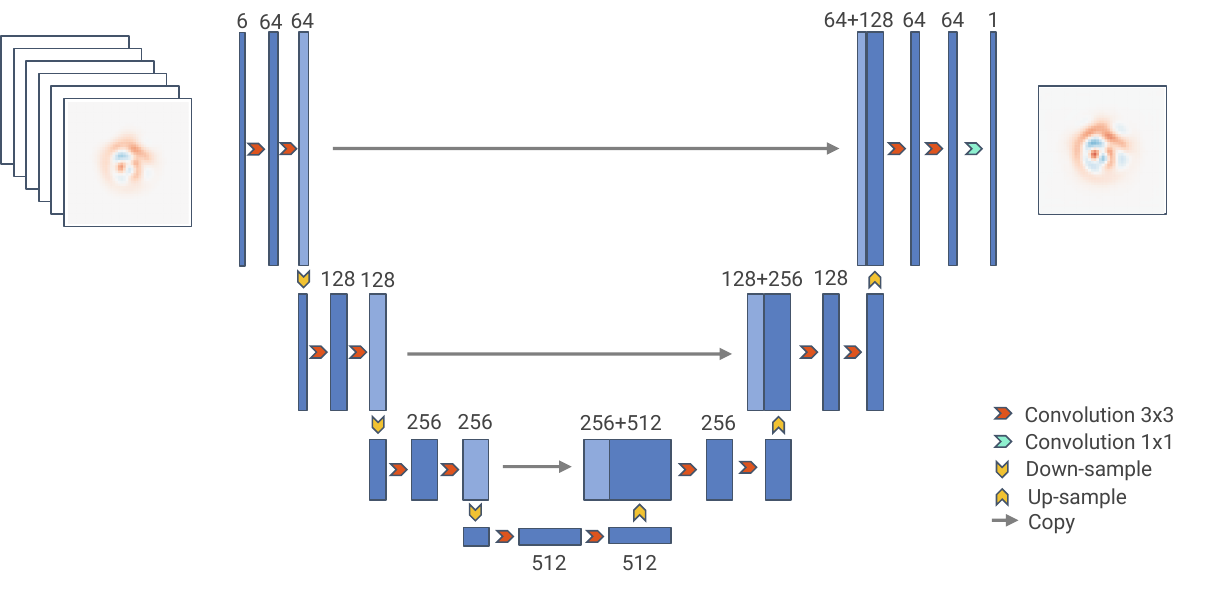}
   
    \caption{Structure of U-Net used for mitigating errors in continuous-variable processes.
    }
    \label{fig:unet}
\end{figure}

\paragraph{U-Net}
The U-Net used in the continuous variable experiment is a modified implementation of the standard U-Net in~\cite{ronneberger2015unet}, as shown in Figure \ref{fig:unet}.
The network contains one embedding block, 3 down-sample blocks, 3 up-sample blocks and 1 output layer. Specifically, one down-sample block contains 2 convolution blocks and 1 MaxPooling~\cite{6144164} operation. Each convolution block is composed by 1 convolution operation~\cite{oshea2015introduction}, followed by 1 LeakyReLU~\cite{maas2013rectifier} and 1 BatchNorm~\cite{ioffe2015batch}. One up-sample block contains 2 convolution blocks and 1 Bilinear interpolation~\cite{gonzales1987digital}. The input to the network is $(g_i, \boldsymbol{P}_i)$. $\boldsymbol{P}_i$ is a tensor of shape $5 \times 48 \times 48$, which corresponds to the Wigner quasi-probability distribution with 5 different noise parameters. Before embedding, $g_i$ is repeated, reshaped into the shape of the Wigner quasi-probability distribution, and stacked with $\boldsymbol{P}_i$ to form a $6\times 48\times 48$ tensor. This tensor is sent into the embedding block of the network, which consists of two convolution blocks. The output layer is one convolution operation with kernel size 1 and Tanh activation. The output of the network $\hat{\boldsymbol{p}}^{(0)}_i$ is the mitigated Wigner function with shape $1 \times 48 \times 48$. We train the network for 300 epochs using Adam optimizer with initial learning rate 2e-4 and batch size 32. The training procedure is the same as in Algorithm \ref{algorithm:train}.

\section{Simulation details}
\subsection{Non-markovian noise model}
For non-markovian noise, we consider the case that each gate inside the circuit is assigned with an independent bath. The interaction between the gate and its bath is formulated by the Spin-Boson model~\cite{10.1093/acprof:oso/9780199213900.001.0001}, which is described by the Hamiltonians
\begin{align}
    H &= H_S + H_B + H_{SB},\\
    H_B &= \sum_k \omega_k b_k^\dagger b_k,\\
    H_{SB} &= \sum_k \sigma_z \otimes (\lambda_k b_k + \lambda_k^* b_k^\dagger),
\end{align}
where $b_k$ is the annihilation operator for mode $k$, and $\omega_k$ is the corresponding energy. The system Hamiltonian $H_S$ is chosen according to the actual quantum gate. The bath can be characterized by the spectral density~\cite{10.1093/acprof:oso/9780199213900.001.0001}, which is defined as
\begin{equation}
    J(\omega) = |\lambda_k|^2 \delta(\omega - \omega_k).
\end{equation}
In our experiments, we consider the continuum bath with spectral density
\begin{equation}
    J(\omega) = \alpha \omega_c^{1-s} \omega^s e^{-\omega / \omega_c},
\end{equation}
where we set $\alpha = 0.001$, $s = 6$, and $\omega_c = 5$. This corresponds to the situation that the system has weak coupling to the bath.

\subsection{Simulation of quantum circuits}
To allow for unified simulation of both Markovian and non-Markovian noise, we implemented a customized simulator. The simulator uses density matrix simulation strategy, and only support $\{R_x, R_z, \text{CNOT}\}$ as basis gates inside a circuit. All gates from an arbitrary circuits are first converted into these basis gates using the transpile function in Qiskit~\cite{Qiskit}. Next, the gates are further transformed into corresponding system Hamiltonians $H_S$ with evolution time associated with the gate parameters. For instance, if the gate inside the circuit is Pauli-X rotation, e.g., $R_x(\pi / 4)$, then the corresponding system Hamiltonian is $H_S = X$, and the evolution time $t \propto \pi / 4$. For CNOT gate, its corresponding Hamiltonian is
\begin{equation}
    H_{\text{CNOT}_{1,2}} = \frac{\pi}{4}(-Z_1 I_2 + Z_1X_2 - I_1X_2),
\end{equation}
as introduced in \cite{PhysRevB.81.134507}, where index 1 denotes the control qubit and index 2 denotes the target qubit. The evolution time is $t_0 \propto 1$. 

For simulation, in noise-free case, the output state of a quantum gate described by $H_S$ is given as{
\begin{equation}
    \rho' = e^{-iH_St}\rho_0e^{iH_St},
\end{equation}
}
where $\rho_0$ is the input state. Whereas in the noisy case, we solve the evolution in interaction picture, and convert back to Schr\"{o}dinger picture~\cite{lidar2020lecture}. The simulation of the evolution of quantum states is done by NumPy~\cite{2020NumPy-Array} and SciPy~\cite{2020SciPy-NMeth} with numerical integrations involving the density matrix.

\subsection{Simulation of spin-system dynamics}
The initial states are ground states of random Ising model, which are generated by the DMRG algorithm~\cite{white1992density}. The algorithm represents the Hamiltonian and the state as Matrix Product Operator (MPO)~\cite{Bridgeman_2017} and Matrix Product State (MPS)~\cite{Bridgeman_2017} respectively. It searches for the ground state by iteratively updating each site of the MPS to the corresponding ground state of the contracted MPO, sweeping from left to right and then from right to left. For a specific site, Lanczos method~\cite{Lanczos:1950zz} is used to find the ground state of the local MPO. In our implementation, the number of DMRG sweeps is set to be 2. The number of iterations of Lanczos method for diagonalizing the local MPO is 2, and the maximum dimension of Krylov space for the diagonalization is 4.

For simulating the dynamics, we use TEBD algorithm~\cite{white2004real, daley2004time}. It evolves the input state by a Hamiltonian using Trotter-Suziki decomposition~\cite{1976CMaPh..51..183S}. The input state is represented as MPS. The original Hamiltonian is decomposed into multiple local Hamiltonians and formulated into MPO, each of which evolves the state by timestep $dt$. In our implementation, we choose timestep $dt = 2$. To restruct the bond dimension of MPS, the maximum singular values is set to be 25. We truncate the singular values smaller than 1e-10.

The noise is added at the end of the simulation, before measurement. This can be done by changing the observable and using the new observable for measurement. For example, consider a quantum state $\rho$ affected by quantum noise $\mathcal E$. This can be expressed using Kraus operators $\{K_i\}$,
\begin{equation}
    \mathcal E(\rho) = \sum_i K_i \rho K_i^\dagger.
\end{equation}
Measure the output by observable $M$, one obtains
\begin{align}
    \tr (M\mathcal E(\rho)) &= \tr(M\sum_i K_i \rho K_i^\dagger)\\
    &= \tr(\sum_i K_i^\dagger M K_i \rho)\\
    &= \tr(\tilde{M} \rho),
\end{align}
where $\tilde{M} = \sum_i K_i^\dagger M K_i$. Thus we can use this new observable to measure $\rho$, leading to the noisy measurement results.

\subsection{Simulation of continuous-variable process}
The initial state of the system is coherent state, defined as
\begin{equation}
    |\alpha \rangle = e^{-\frac{|\alpha|^2}{2}}\sum_{n = 0}^\infty \frac{\alpha^n}{\sqrt{n!}}|n\rangle,
\end{equation}
which is expanded in the basis of Fock states $\{|n\rangle \}$.
In our experiment, we choose $\alpha = 1.5$ and truncate the number of Fock states in Hilbert space to be $N = 15$.
We simulate the continuous-variable process evolution using QuTiP package~\cite{JOHANSSON20121760}. We consider the Wigner function~\cite{PhysRevLett.78.2547,PhysRevLett.89.200402} of the state as measurement output, which is a 2-dimensional quasiprobability distribution. We quantize the distribution into $48\times 48$ grids, which is enough for describing the state.

\section{Implementation of error mitigation algorithms for comparative experiments}

\subsection{Zero-noise extrapolation}
In our experiments, we use a quadratic function as extrapolation model to fit the noisy measurement data. The data is the same as in our DAEM's input. Note that ZNE requires the exact noise parameters as input, so we provide this additional information, which is not necessary for our neural model. For each different cirucit structure, observable, and initial state, we fit an individual ZNE model using the corresponding measurement data, and extrapolate to zero noise to obtain the mitigated result.

\subsection{Clifford data regression}
Clifford data regression in our experiments is implemented by fitting a linear model with random Clifford data. Specifically, for a target circuit and target observable to be mitigated, we generate 100 different Clifford circuits by replacing the single-qubit gates in the target circuit with random Clifford gates, while the CNOT gates are left unchanged. The Clifford data is generated by measuring the noisy Clifford circuits using the target observables, and classically simulating the corresponding noise-free measurements. We train the linear model with the generated Clifford data until convergence. Then we apply the trained model to mitigate errors. Note that for each different circuit and observable, individual Clifford data is generated to train the model.

\section{Experiments on scalability of DAEM}
\begin{figure}[h]
    \centering
    \includegraphics[width=0.35\linewidth]{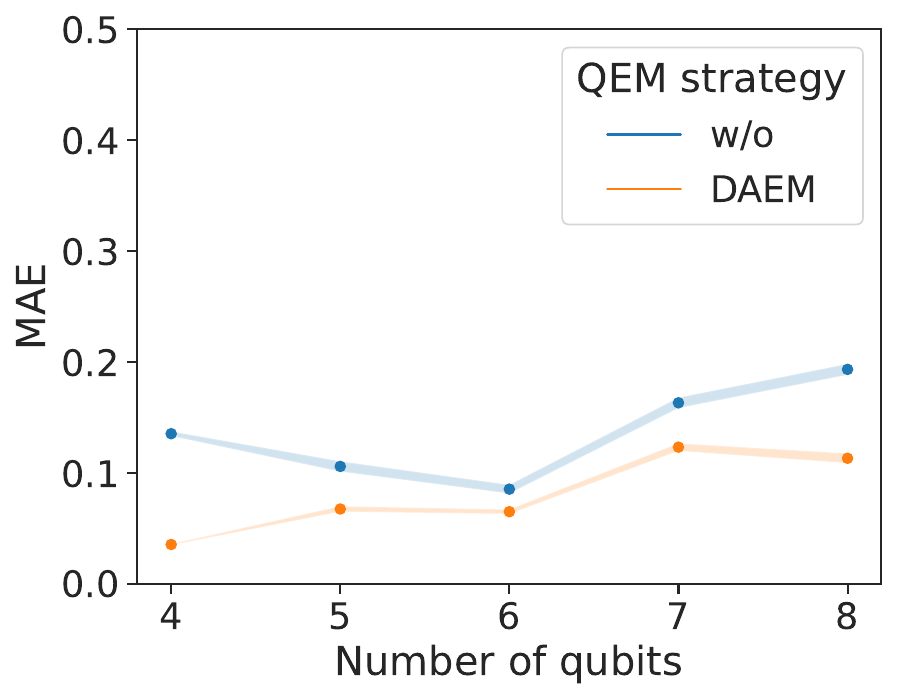}
    \caption{Performance of DAEM with respect to different number of qubits.}
    \label{fig:vqe_numqubits}
\end{figure}

\begin{figure}[h]
    \centering
    \includegraphics[width=0.35\linewidth]{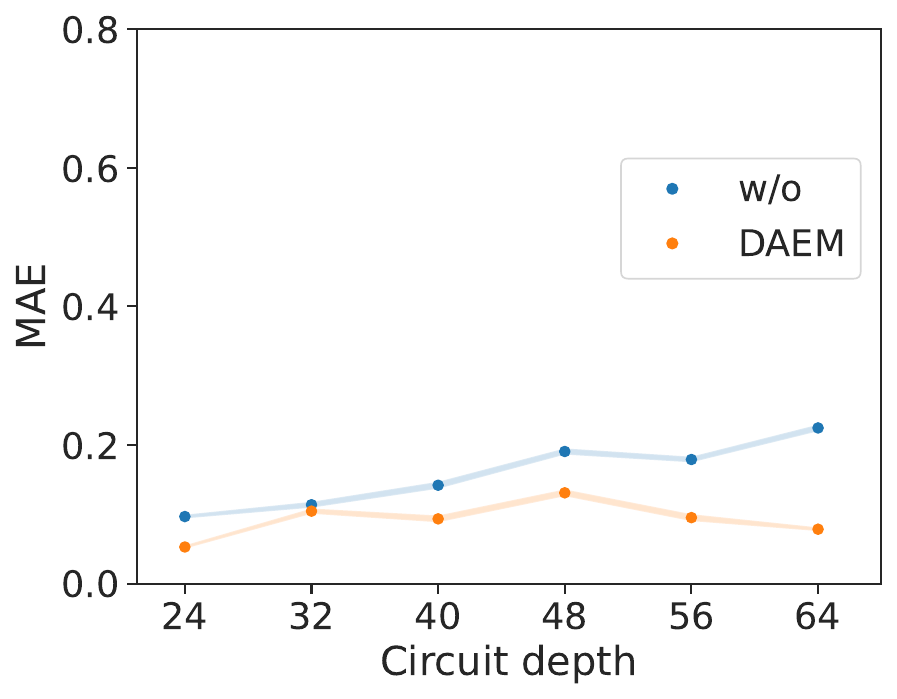}
    \caption{Performance of DAEM with respect to different circuit depth.}
    \label{fig:vqe_depth}
\end{figure}
{
We examine the scalability of our model with respect to circuits of different numbers of qubits and different circuit depth. For all the following experiments, we fix the number of initial states of the fiducial circuits to 150. For each different qubit number and circuit depth, we generate new fiducial circuits for training. The trained model is executed on 10 different trained VQE circuits, in which the coefficients of Ising model $g\in [1.0, 2.0]$ with stride 0.1. The noise model is chosen as phase damping noise.

To test the performance with respect to different numbers of qubits, we fix the circuit size to 9, and change the number of qubits from 4 to 8. The results in Figure \ref{fig:vqe_numqubits} show that our model is potentially scalable to larger systems, as the performance remains stable despite the fixed training dataset size. In addition, in the Hamiltonian evolution experiment with a system size up to 50 qubits, our model also demonstrates good performance using a training set of 100 different states, further verifying the scalability of our model.

To showcase the performance of our model with respect to circuit depth, we fix the number of qubits to 4, and vary the depth of the circuit from 24 to 64. Results in Figure \ref{fig:vqe_depth} show that without mitigation, the gap between noisy and ideal measurement expectation values increase with the circuit depth, but with our mitigation method, the gap keeps stable.

}

\end{document}